\documentclass{emulateapj}

\newcommand{\be}{  \begin{eqnarray} }
\newcommand{\ee}{  \end{eqnarray} }
\newcommand{\bd}{  \begin{displaymath} }
\newcommand{\ed}{  \end{displaymath} }

\newcommand{\msun}{ M_{\odot}}
\newcommand{\mdot}{\dot{M}}
\newcommand{\vmg}{v_{\rm Mg}}
\newcommand{\mvir}{M_{\rm vir}}
\newcommand{\teff}{T_{\rm eff}}
\newcommand{\tin}{T_{\rm in}}
\newcommand{\kb}{k_{\rm B}}

\begin{document}

\title{The UV Continuum of Quasars: Models and SDSS Spectral Slopes} 

\author{Shane W. Davis\altaffilmark{1,2,3}, 
Jong-Hak Woo\altaffilmark{3}, and 
Omer M. Blaes\altaffilmark{3}}

\altaffiltext{1}{Institute for Advanced Study, Einstein Drive, 
Princeton, NJ 08540}
\altaffiltext{2}{Chandra Fellow}
\altaffiltext{3}{Department of Physics, University of California, 
Santa Barbara, CA 93106}

\begin{abstract}

We measure long (2200-4000 \AA) and short (1450-2200 \AA) wavelength
spectral slopes $\alpha$ ($F_{\nu} \propto \nu^{\alpha}$) for quasar
spectra from the Sloan Digital Sky Survey.  The long and short
wavelength slopes are computed from 3646 and 2706 quasars with
redshifts in the z=0.76-1.26 and z=1.67-2.07 ranges, respectively. We
calculate mean slopes after binning the data by monochromatic
luminosity at 2200 \AA\ and virial mass estimates based on
measurements of the Mg~II line width and 3000 \AA\ continuum
luminosity.  We find little evidence for mass dependent variations in
the mean slopes, but a significant luminosity dependent trend in the
near UV spectral slopes is observed with larger (bluer) slopes at
higher luminosities.  The far UV slopes show no clear variation with
luminosity and are generally lower (redder) than the near UV slopes at
comparable luminosities, suggesting a slightly concave quasar
continuum shape.  We compare these results with Monte Carlo
distributions of slopes computed from models of thin accretion disks,
accounting for uncertainties in the mass estimates.  The model slopes
produce mass dependent trends which are larger than observed, though
this conclusion is sensitive to the assumed uncertainties in the mass
estimates.  The model slopes are also generally bluer than observed,
and we argue that reddening by dust intrinsic to the source or host
galaxy may account for much of the discrepancy.

\end{abstract}


\section{Introduction}
\label{intro}

The ``bare'' thin accretion disk has been invoked ubiquitously to
explain both general and detailed properties of a wide variety of
accreting systems.  In systems which are believed to harbor black
holes the results have been mixed.  The timing and spectral variations
of Galactic black hole candidates show a range of behavior, much of
which cannot be accommodated by a simple thin accretion disk.  However,
some of these sources enter a thermal state, where the spectral energy
density (SED) is dominated by emission which can be well fit with a
simple multitemperature blackbody model \citep{mit84} or
more elaborate variations \citep[e.g.][]{li05,dah06}.

The thin disk model may also explain a portion of the emission in
Active Galactic Nuclei (AGNs).  These sources are almost certainly
powered by an accretion flows onto a central ``super-massive'' black
hole \citep{kro99}.  As a result, it is often supposed that the broad
UV peak (``big blue bump'') of the SED is emission from a thin
accretion disk.  However, this interpretation faces a number of
challenges when detailed comparisons are made between models and the
data \citep[see e.g.][]{kab99}. It may be possible to resolve some of
these discrepancies by modifying the model and considering additional
processes (e.g. dust reddening, irradiation, inhomogeneities, and
multi-phase flows), but it is then pertinent to ask what, if any,
predictive power is provided by the thin disk model itself.

Models of thin accretion disks predict that the effective temperature
at the inner-most radius $T_{\rm in}$ of the disk will be proportional
to the one-fourth power of accretion rate $\mdot$ and inversely
proportional to one-half power of black hole mass $M$ ($T_{\rm in}^4
\propto \mdot/M^2$).  This scaling follows rather generally from the
assumptions that gravitational binding energy is radiated locally and
that the relevant length scale of the emitting region is the
gravitational radius $R_g=G M/c^2$ of the black hole.  As we shall
discuss, this relation accounts for much (but not all) of the spectral
dependence on $M$ and $\mdot$ since $T_{\rm in}$ roughly determines
the photon energy where the SED peaks ($\sim k_{\rm B} T_{\rm in}$).

In fact, one of the successes of this scenario is that it
approximately predicts the position of the continuum peak for both
super-massive black holes in AGN and the $\sim 10 \msun$ black holes in
Galactic X-ray binaries, assuming that both sources accrete mass at
similar fraction of the Eddington limit. Also, the approximate
relation $T_{\rm in}^4 \propto L \propto \mdot$ (where $L$ is the
bolometric luminosity) can be inferred from the spectral evolution of
several black hole binaries in thermal state \citep[see
e.g.][]{gad04}.  Unfortunately the narrow range of dynamically
inferred masses, uncertainties in the estimates, and small sample of
sources complicate efforts to simultaneously and independently
constrain the $M$ dependence of the SED in these systems.

In many respects, AGNs offer greater promise for testing the $M$ and
$L$ (or $\mdot$) dependence of thin disk spectral models since there
is a much larger sample available which spans a wider range of $M$ and
$L$.  However, there are also a number of additional
challenges.  First, mass estimates in AGN are generally more uncertain
than in black hole candidates, and the most reliable methods can only
be applied in a small fraction of sources.  Also, the SED peaks in the
far UV so it is challenging to get broadband coverage at low
redshifts, except for a relatively small number of bright sources.
Finally, the large number of emission lines which characterize most
AGN may prevent a robust determination of underlying continuum
emission.

Despite these challenges, comparing the predicted and observed
spectral evolution as a function of $M$ and $L$ is the principle aim
of this work.  To accomplish this, we measure UV spectral slopes
($\alpha$ where $F_\nu \propto \nu^\alpha$) for several thousand
quasars from the Sloan Digital Sky Survey (SDSS).  A wide range of
$\alpha$ is inferred, allowing us to examine the extent to which
$\alpha$ correlates with the $M$ and $L$, the parameters which
determine the model SEDs \citep[see e.g.][]{sam89,lan89,hub00}.  We
compare the observed slopes with calculations based on the relativistic,
fully non-LTE models of \citet{hub00}.  In addition to allowing us to
test the predictions of the thin disk model, a correlation (or lack
thereof) may provide clues and place important constraints on other
processes which may be important for determining the continuum SED.

The plan of this work is as follows. In order to test the predictions
of the thin disk model we construct a large table of artificial SEDs
for direct comparison with data.  The methods used to construct these
models are summarized in \S\ref{models}.  In \S\ref{mass} we briefly
review the mass estimation methods employed in this work as our
conclusions are sensitive to the reliability of these estimates.  In
\S\ref{fuse} we compare the model SEDs with a small sample of
well-observed, relatively nearby AGNs with simultaneous optical to UV
spectra \citep{sha05}. In \S\ref{sdss} we present slope measurements
for a large sample of SDSS QSOs.  In \S\ref{mcmod} we calculate slopes
from our spectral models and generate Monte Carlo realizations of the
slope distributions for comparison with the data.  In \S\ref{discus}
we discuss the possible origins of discrepancies between the models
and observations, considering additional processes, not accounted for
by the models, which might be important.  In \S\ref{comp}, we compare
our results with previous work, particularly that of \citet{bon06}
which is the most similar to our current efforts. We summarize our
conclusions in \S\ref{conc}.  Throughout this work, we use the
following cosmology: $H_0=70 \;{\rm km \; s^{-1} \; Mpc^{-1}}$,
$\Omega_m=0.3$, and $\Omega_{\Lambda}=0.7$.

\section{Spectral Models}
\label{models}

Our artificial SEDs are based on time-independent models of thin,
$\alpha$-disks \citep{sas73}.  Here, $\alpha$ refers to an assumed
constant of proportionality between the accretion stress and total
pressure. Therefore, it is a dimensionless parameter and generally
assumed to be less than unity.  (Hereafter, we will use $\alpha_{\rm
SS}$ to differentiate this quantity and the spectral slope $\alpha$,
but we will continue to refer to the model simply as an
$\alpha$-disk.)  We generate artificial SEDs using AGNSPEC, an
interpolation scheme which is identical to that described in
\cite{dah06}.  Except for the interpolation scheme, the models used
here are equivalent to those presented in \citet{hub00}.  We only
summarize the most relevant features and the reader is referred to
this work (and references therein) for a more detailed discussion.

The artificial SEDs are based on a relativistic, thin, $\alpha$-disk
model similar to that of \citet{nat73}, but include some minor
corrections \citep{rah95}.  These SEDs account for the effects of
light bending and time dilation by calculating the null geodesics of
the black hole spacetime \citep[KERRTRANS,][]{ago97}.  In addition to
$M$ and $\mdot$, such fully-relativistic models require a choice of
the black hole spin parameter $a_\ast=cJ/G M^2$, where $J$ is the
angular momentum of the hole.

Using these ``one-zone'' disk models as a basis, we compute two types
of artificial SEDs.  We first construct relatively simple spectra
which assume the disk surface emits like a blackbody at the local
effective temperature (but still account for relativistic effects).  A
second set of more sophisticated models employ TLUSTY \citep{hal95} to
solve the coupled equations of radiative transfer and equilibrium,
non-LTE vertical structure in the disk.  In this case the disk surface
density is needed to determine the structure, requiring a model for
angular momentum transport.  We adopt an $\alpha$-disk prescription
with $\alpha_{\rm SS}=0.01$.  At the wavelengths of interest the
spectra are relatively insensitive to $\alpha_{\rm SS}$ for the ranges
of parameters ($a_\ast$, $M$, and $L$) explored here.  These models
include bound-free and free-free opacities of H and He and the effects
of electron scattering are calculated in the Thomson limit.  Due to
the difficulties involved, we do not include opacity from bound-bound
transitions, although it may have a significant effect on the
resulting spectra \citep[see e.g.][]{hah98}.

The inclusion of non-LTE effects and realistic opacities can cause
significant shifts from a simple blackbody spectrum.  The effects are
usually largest at frequencies near the bound-free transitions.  Since
we concentrate on the SED at wavelengths longer than 1000 \AA, the
Balmer edge will be the most important.  Unfortunately,
the annuli which produce most of the flux longward of the Balmer edge
are the most difficult to calculate due the presence of ionization zones
associate with the transition.  In these annuli, equilibrium solutions
are either unattainable or lead to unstable atmospheres with density
inversions \citep[see \S 3.3 of][]{hub00}.  Due to these difficulties,
we simply assume blackbody spectra for annuli with effective temperatures
below 9,000 K.  Based on Fig. 11 of \cite{hub00}, this probably leads to
a slight underestimate of flux at 4000 \AA\ and possible implications
are discussed in \S\ref{lumdep}.

\section{Methods}
\subsection{Mass Estimates}
\label{mass}

\begin{deluxetable*}{lcccccccccc}
\tablecolumns{11}
\tablecaption{Slope Comparison \label{tbl1}}
\tablewidth{0pt}
\tablehead{
\colhead{Source} &
\colhead{Mass\tablenotemark{a}} &
\multicolumn{3}{c}{$\alpha (1450-2200 {\rm \AA})$} &
\multicolumn{3}{c}{$\alpha (2200-4000 {\rm\AA})$} &
\multicolumn{3}{c}{$\alpha (1450-4000 {\rm\AA})$} \\
 &
\colhead{($10^8 \msun$)} &
\colhead{Data} &
\colhead{$a=0$} &
\colhead{$a=0.9$} &
\colhead{Data} &
\colhead{$a=0$} &
\colhead{$a=0.9$} &
\colhead{Data} &
\colhead{$a=0$} &
\colhead{$a=0.9$} \\
}

\startdata
3C 273 &
8.86 &
-0.64 &
-0.34 &
-0.24 &
0.08 &
-0.16 &
-0.11 &
-0.21 &
-0.23 &
-0.16 \\
PG 0953+414 &
2.76 &
-0.16 &
-0.17 &
-0.12 &
-0.27 &
-0.12 &
-0.080 &
-0.23 &
-0.14 &
-0.095 \\
PG 0052+251 &
3.69 &
-0.60 &
-0.21 &
-0.038 &
-0.28 &
-0.28 &
-0.14 &
-0.41 &
-0.25 &
-0.097 \\
Ton 951 &
0.92 &
-0.98 &
0.00 &
0.073 &
-0.57 &
-0.14 &
-0.074 &
-0.74 &
-0.08 &
-0.014 \\
Mrk 509 &
1.43 &
-1.0 &
-0.074 &
0.092 &
-0.19 &
-0.24 &
-0.096 &
-0.52 &
-0.17 &
-0.019 \\

\enddata
\tablenotetext{a}{Reverberation mapping estimates from \citet{pet04}.}

\end{deluxetable*}

Stellar dynamical estimates of black hole masses are only available in
the local universe.  At higher redshifts virial mass estimates based
on reverberation mapping of broad line region (BLR) clouds \citep[see
e.g.][]{pet04} are generally accepted to be the most reliable
\citep[see, however,][]{kro01}.  This technique uses the
full-width-at-half-maximum (FWHM) or second moment of one or more
prominent broad emission lines to estimate the velocity field $v_{\rm
BLR}$ of broad line clouds.  Reverberation mapping then provides a
characteristic radius $R_{\rm BLR}$ from which the virial mass $\mvir
\sim v_{\rm BLR}^2 R_{\rm BLR}/G$ can be estimated. The precise
normalization depends on the kinematics of the BLR and cannot, in
general, be determined reliably for individual sources.  A single
normalization for all sources can be obtained by requiring virial
estimates to lie on the $M-\sigma$ relation \citep{onk04}.

A significant drawback of this method is that it requires sources to
be frequently monitored and can only be robustly applied in cases
where the lines provide a clear response to continuum variations.  As
a result, it has only been used successfully for relatively nearby
sources. \citep{wpm99,kas00,kas05}.  At higher redshifts, empirically
calibrated luminosity--radius relations are more commonly used
\citep[e.g.][]{ves02,wau02,vap06}.  For
these methods it is generally assumed that $R_{\rm BLR} \propto
(\lambda L_\lambda)^\delta$ where monochromatic luminosity $\lambda
L_\lambda$ is calculated at some particular rest wavelength.  This is
usually chosen to be 5100 \AA\ when $v_{\rm BLR}$ is estimated from
the width of H$\beta$. The constant of proportionality and exponent
$\delta$ are then fit to minimize the scatter between this relation
and estimates from a reverberation mapped sample.  Typically $\delta
\sim 0.5$ \citep[see e.g.][]{ben06} is obtained, consistent with
expectations from photoionization models of the BLR.

At higher redshifts, H$\beta$ shifts out of the observed band and
other broad emission lines are used, typically Mg~II or C~IV.  The
estimates used in \S\ref{sdss} were calculated from line width
measurements obtained by \citet{mad04} and the reader is referred to
their paper for a detailed discussion.  Since we are only considering
quasars with $z \gtrsim 0.7$ all estimates utilize the FWHM of Mg~II
$\vmg$ and $L_{3000}$ the monochromatic luminosity measured at 3000
\AA.  Therefore, the mass estimates are calculated using equation (A6)
of \citet{mad04}
\be
\frac{\mvir}{\msun}=3.2 \left(\frac{L_{3000}}{10^{44} {\rm \; erg
\;s^{-1}}}\right)^\delta \left(\frac{\vmg} {\rm km \; s^{-1}}
\right)^2.  \label{eq:mass} 
\ee 
Using $L_{3000}$, \citet{maj02} obtain $\delta =0.47$ by fitting the
$R_{\rm BLR}-L_{3000}$ relation to best match reverberation mapped
masses of 34 AGNs \citep{kas00}.  However, this relation is derived
including a number of relatively low luminosity AGNs while the $z >
0.7$ SDSS quasar sample only includes sources with $L_{3000} > 10^{44}
\rm \; erg \; s^{-1}$.  Therefore, \citet{mad04} refit the $R_{\rm
BLR}-L_{3000}$ relation only using those sources in the reverberation
mapped sample with $L_{3000} > 10^{44} \rm \; erg \; s^{-1}$, finding
$\delta = 0.62$.  Since our sample is a subset of the \citet{mad04}
sample which only includes this luminosity range, we also adopt the
$\delta=0.62$ value.

\begin{figure*}
\plotone{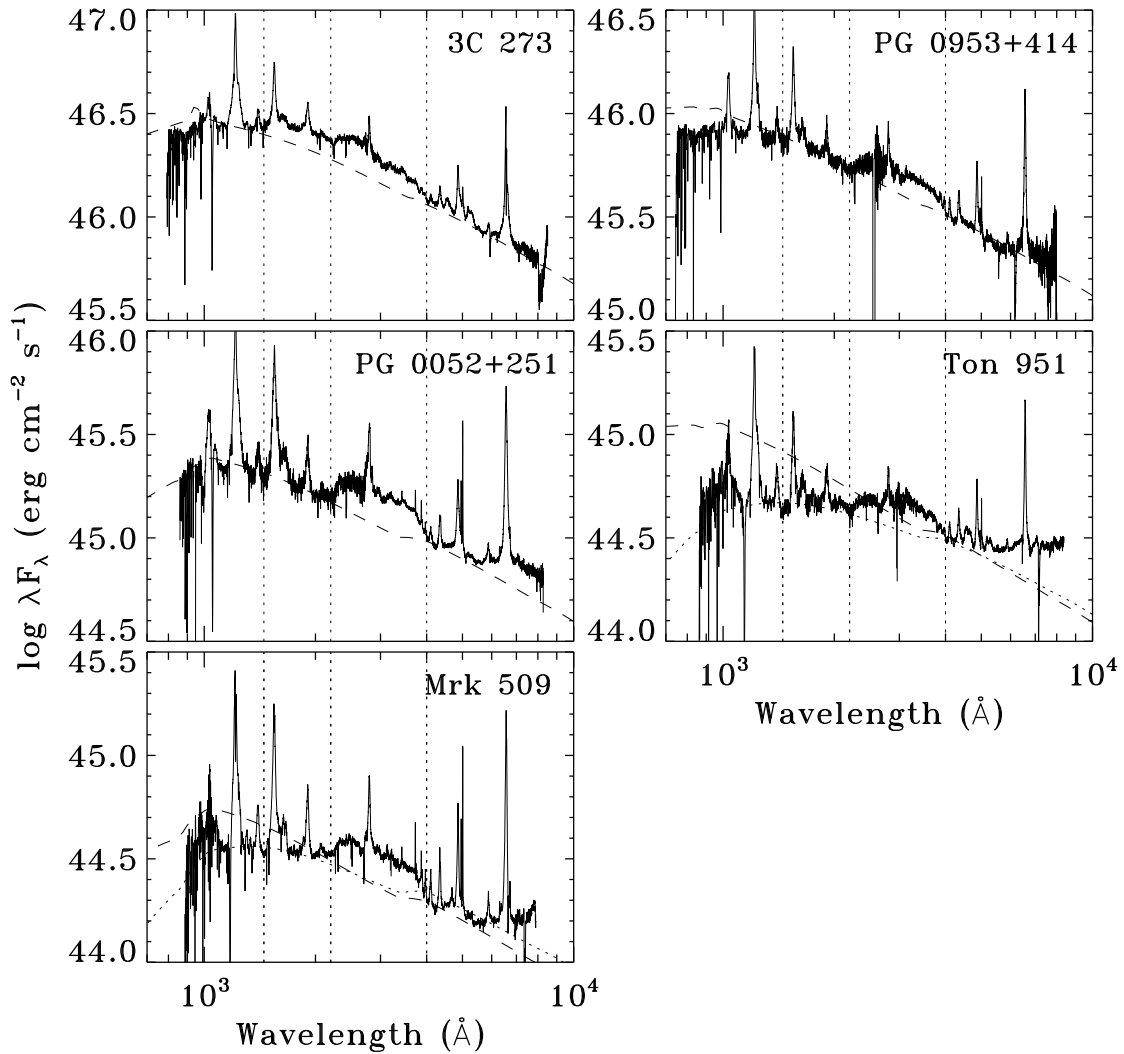}
\caption{Optical to UV SEDs of five AGN from the sample of
\citet{sha05}.  The dashed curves are model SEDs with $M$
corresponding to the the reverberation mapping estimates in Table
\ref{tbl1} and $a_\ast=0$.  We choose the inclinations $i$ so that
$\cos i = 0.8$ for all models except 3C 273.  In this case we use
$\cos i = 1$ since the source exhibits superluminal motion.  The
accretion rate is fixed by approximately matching the model and data
at 4000 \AA.  In Ton 951 and Mrk 509 the bare models are clearly a
poor approximation so we also consider models (dotted covers) which
are reddened using an SMC-like extinction curve. See the text in
\S\ref{fuse} for further discussion.
\label{fig:fuse}}
\end{figure*}

\subsection{Comparing Models to Data with Spectral Slopes}
\label{fuse}

One of the most striking characteristics of most AGNs (particularly
quasars and type I Seyferts) are the numerous strong broad, emission
lines.  A substantial fraction of the UV and optical flux can be
emitted in these lines, making a precise and robust identification of
the continuum difficult, if not impossible.  In some cases the broad
line region emission appears to be unpolarized and the underlying
continuum can be identified with polarimetry \citep{kab03,kis04},
revealing a Balmer edge beneath the ``small blue bump''.  Although
promising, this technique requires polarimetry observations and
unpolarized BLR emission, making it unsuitable for our purposes.
Alternatively, we could attempt to model the BLR emission.
Photoionization models can qualitatively reproduce many aspects of the
observed spectra, but detailed fits to individual SEDs remain a
challenge.  Therefore, we feel the substantial work involved in
generating such models and fitting the data is best left for future
efforts.

Lacking a suitable method for removing the BLR emission, we simply
adopt ``continuum windows'' which appear to be largely devoid of BLR
contamination and focus on the flux at these wavelength for comparison
with the models.  Unfortunately, no part of the spectrum is completely
devoid of line emission, and the best one can hope to do is choose
regions which are relatively free of BLR contamination.  Our choices
are guided in part by the SDSS quasar composite of \citet{van01}.  By
constructing a composite spectrum from thousands of quasars, one can
construct a high signal-to-noise SED which is sensitive to the
presence of rather weak emission features.  We focus on three windows
near 1450 \AA, 2200 \AA, and 4000 \AA.  

In making these choices, we have attempted to balance the requirement
of little contamination with a desire to have as large a sample of
quasar spectra as possible which cover two of the windows.  Of these
choices, 2200 \AA\ is probably the most problematic due to the
presence of Fe~II contamination \citep[see e.g. Fig. 6 of][]{van01}.
However, there are no preferable options between 1450 \AA\ and 4000
\AA.  The SDSS spectra cover an observed wavelength range from 3800
\AA\ to 9200 \AA, roughly a factor 2.4.  Therefore, 1450 \AA\ and 4000
\AA\ cannot be simultaneously covered with SDSS, making an
intermediate window necessary.

Although we do not have full UV spectral coverage for the SDSS
quasars, we can gain insight by first comparing our models with
broadband spectra of lower redshift AGN.  The sample of
\citet{sha05} is ideally suited for this purpose.  It
combines nearly simultaneous spectra for 17 relatively bright, nearby
AGN taken with {\it Far Ultraviolet Spectroscopic Explorer}, {\it
Hubble Space Telescope}, and the 2.1 m Kitt Peak National Observatory.
\citet{sha05} provide a detailed analysis of the sample, including
comparisons with the spectral models described in \S\ref{models}.
Here we only provide a brief comparison of five sources with
reverberation mapping mass estimates \citep{pet04}.

As shown in Figure \ref{fig:fuse}, the SEDs of these sources were
shifted to rest wavelength and corrected for Galactic reddening
using the values from Table 1 of \citet{sha05} which were obtained
from NED and based on the \citet{she98} map. The centers of the
continuum windows are marked with dashed vertical lines.  For
comparison we plot thin disk models for $a_\ast=0$ black holes as
dashed curves.  The models can be very roughly approximated by
power laws except for the breaks near the Lyman and Balmer edges at 912
\AA\ and 3650 \AA.  The broad feature between 2200 \AA\ and 4000 \AA\
(``small blue bump'') is predominantly a superposition of Balmer
continuum and a plethora of Fe~II emission lines produced by the BLR,
and, therefore, not accounted for by the model SEDs.

The black hole masses used for the models are equivalent to the
reverberation mapping estimates which are listed in the second column
of Table \ref{tbl1} \citep{pet04}.  The model $\mdot$ is chosen to
approximately match the observed flux at 4000 \AA, assuming an
inclination $i$ such that $\cos i = 0.8$ for all cases except 3C 273.
In the case of 3C 273, we used $i=0$ since superluminal motion is
observed in this source.  The model bolometric luminosities in
Eddington units are 0.32, 0.63, 0.063, 0.16, and 0.035 for 3C 273, PG
0953+414, PG 0052+251, Ton 951, and Mrk 509, respectively.

Qualitatively, the comparisons provide mixed results.  They are
somewhat better for the higher luminosity AGN, but the the model
spectra are clearly too steep in the UV for the lowest luminosity
sources (Ton 951 and Mrk 509).  PG 0953+414 and PG 0052+251 yield the
best results in 1450-4000 \AA\ band, although they respectively
overpredict and underpredict the emission below 1000 \AA.  Although
the model provides an approximate reproduction of the continuum in 3C
273 for $\lambda \lesssim 1450$ \AA\ and $\lambda \gtrsim 4000$ \AA,
it clearly underestimates the flux at 2200 \AA.  In PG 0052+251, Ton
951, and Mrk 509, the models clearly underestimate optical continuum.
We attribute some fraction of this discrepancy to contamination from
the host galaxy.  This conclusion is supported by the apparent
anticorrelation between the optical ``excess'' and AGN luminosity as
the fractional contamination should be weaker in the more luminous AGN
if we assume an approximately constant host contribution.

Intrinsic reddening due to dust in the host galaxy may contribute to
some of the discrepancy between data and models in Ton 951 and Mrk
509.  However, there is disagreement about the form of the reddening
curve in QSOs, an issue we discuss further in \S\ref{dust}.  As an
example, we plot models (dotted curves) which have been
reddened with the ``SMC-like'' reddening curve \citep{pre84}
$A_\lambda=1.39 \; E(B-V) (\lambda/\mu {\rm m})^{-1.2}$ adopted by
\citet{ric03}.  We adjust E(B-V) and $\mdot$ in an attempt to obtain
good agreement at 1450 \AA\ and 2200 \AA, but still require the flux
to match the observed value at 4000 \AA.  We find best agreement for
$E(B-V)\sim 0.055$ and 0.04 for Ton 951 and Mrk 509, respectively.
Although this improves the agreement in the longward of $\sim 1000$
\AA, the models underestimate the flux at short wavelengths.  We note
that reddening with this extinction curve can only worsen the ``fit''
to 3C 273, since the flux at 2200 \AA\ is already {\it above} the
model prediction.

Table \ref{tbl1} provides a more quantitative comparison using
spectral slopes.  We calculate spectral slopes $\alpha$ ($F_\nu
\propto \nu^\alpha$) defined by
\be 
\alpha \equiv -\left(2+\frac{\log F_\lambda(\lambda_{\rm max}) - \log
F_\lambda(\lambda_{\rm min})}{\log \lambda_{\rm max}- \log
\lambda_{\rm min}} \right).  \label{eq:slope}
\ee
For the three combinations of continuum windows (1450-2200 \AA,
2200-4000 \AA, and 1450-4000 \AA) in Table \ref{tbl1}, we tabulate
three different slopes: one from the data and two for models with
$a_\ast=0$ and 0.9.  The $a_\ast=0$ slopes correspond to the models
plotted in Figure \ref{fig:fuse}.  The $a_\ast=0.9$ models are
selected using the same comparison criterion as described above for
the $a_\ast=0$ models.  The $a_\ast=0.9$ slopes are bluer (greater or
less negative $\alpha$) than the $a_\ast=0$ slopes
(cf. Fig. \ref{fig:model}), and generally provide a poorer match to
the observed SEDs.  For the best case, PG 0953+414, the the model
slopes differ from the continuum by as much $\Delta \alpha \sim 0.15$
for $a_\ast=0$ and $\Delta \alpha \sim 0.2$ for $a_\ast=0.9$.  For the
worst case, Mrk 509, the disagreement can be as high as $\Delta \alpha
\gtrsim 1$.  

These comparisons suggest that measurement of two slopes (between
1450-2200 \AA\ and 2200-4000 \AA) can provide a sensible
parameterization of the UV continuum.  They also
indicate how problems might arise.  Due to the small baselines used
($\Delta \log \lambda =0.18$ and 0.26), a relatively modest difference
in $F_\lambda$ can produce substantial discrepancy in the slope.  For
example, a relative increase of 10\% in $F_\lambda$ at 2200 \AA\ would
yield $\Delta \log \alpha = -0.23$ and $+0.16$, for the 1450-2200 \AA\
and 2200-4000 \AA, respectively.  This makes the observed slopes
particularly sensitive to any additional emission which may
contaminate the continuum windows.  A related concern is that a modest
amount of dust reddening \citep[cf.][]{ric03} in the source or host
galaxy can substantially alter the observed $\alpha$.  For example,
$E(B-V)=0.03$ with the above reddening curved leads to $\Delta
\alpha = 0.37$ and 0.2 for 1450-2200 \AA\ and 2200-4000 \AA\ ranges,
respectively.  Therefore, we consider the effects of reddening in
\S\ref{dust}.

\section{Spectral Slopes of SDSS Quasars}
\label{sdss}

\subsection{Sample and Data Reduction}
\label{sample}

A principle aim of this work is to look for correlations in the
continuum of SDSS quasar spectra with $M$ and $\mdot$.  For our
purposes $\mdot$ can be estimated from the observed luminosity, and
$\mvir$ can be inferred from a combination of $v_{\rm Mg}$ and
$L_{3000}$ as discussed in \S\ref{mass}.  These quantities have
already been computed by \citet{mad04} for a large sample of sources
from the SDSS Quasar Catalogue II \citep{sch03}.  Therefore, we
restrict our attention to this sample and refer the reader to
\citet{mad04} for further details.

From these spectra we must further restrict ourselves to two subsamples
discriminated by their redshift.  At longer wavelengths, our slope
estimates are calculated using the flux measured at $\lambda=2200$ and
4000 \AA.  The SDSS spectra cover an observed wavelength range from
3800-9200 \AA, so the requirement to observe both wavelengths
simultaneously restricts us to a range of $0.76 \lesssim z \lesssim
1.26$.  For the shorter wavelength slope, we need coverage from
$\lambda=1450 - 3000$ \AA\ in order to calculate the slope and the
value of $L_{3000}$ which is required for the mass estimate.  This
restricts our second subsample to $1.67 \lesssim z \lesssim 2.09$.
After applying these cuts we are left with 3783 and 2757 spectra for
the low and high $z$ samples, respectively.

The measurements of $v_{\rm Mg}$ and $L_{3000}$ in \citet{mad04} were
performed using spectra from the SDSS First Data Release \citep{aba03}
while our spectra were obtained from SDSS Fourth Data Release
\citep[DR4,][]{ade06}.  A recalculation of $v_{\rm Mg}$ is beyond the
scope of this work, but we have recomputed $L_{3000}$ with DR4 data,
and confirmed that the modest differences in $L_{3000}$ have little
impact on the inferred mass distribution.

\begin{figure}
\includegraphics[width=0.46\textwidth]{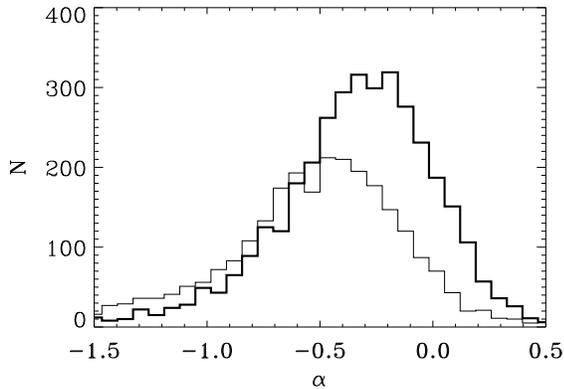}
\caption{Histograms of $\alpha$ calculated for two samples of SDSS
QSOs.  The thin solid curve corresponds the distribution of $\alpha$
measured from 1450-2200 \AA\ for a sample of QSOs with $1.67 \lesssim
z \lesssim 2.09$.  The thick solid curve corresponds the distribution
of $\alpha$ measured from 2200-4000 \AA\ for a sample of QSOs with
$0.76 \lesssim z \lesssim 1.26$.
\label{fig:slopedist}}
\end{figure}

\subsection{Spectral Slopes}
\label{slopes}

\begin{figure*}
\plotone{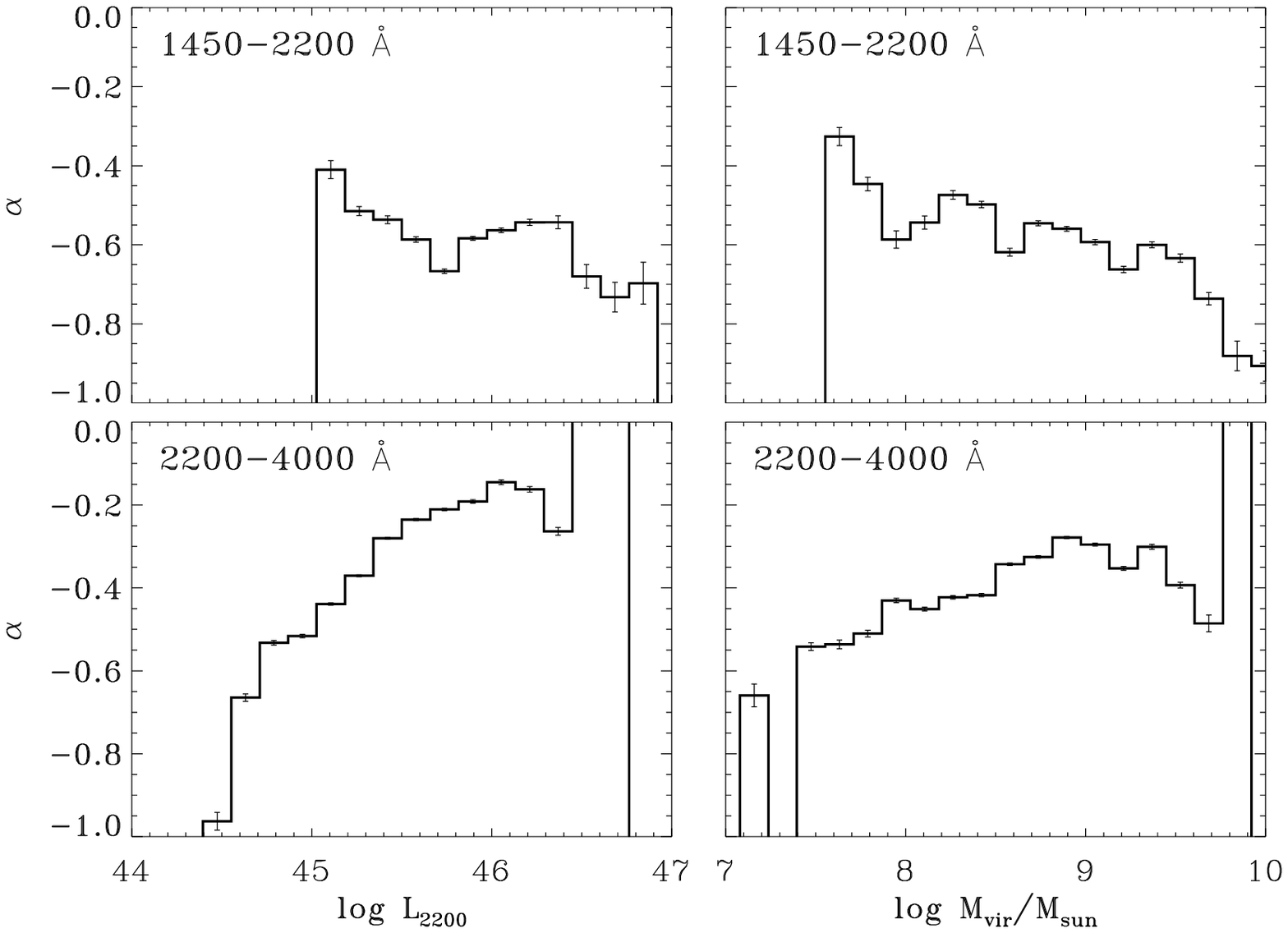}
\caption{1D distribution of mean slopes $\alpha$ binned by
monochromatic luminosity at 2200 \AA\ (left) and viral mass (right).
The top and bottom panels correspond to slopes measured from 1450-2200
\AA\ and 2200-4000 \AA, respectively. Error bars are calculated
assuming uncertainties $\sigma/\sqrt{N}$ where $\sigma$ and $N$ are
the standard deviation and number of spectra in each bin.
\label{fig:1d}}
\end{figure*}

In order to compute $\alpha$, we must extract the flux at the
wavelengths of interest.  We first correct for Galactic reddening
using the \citet{she98} map, and then transform to
the quasar rest frame.  Fluxes and their uncertainties are then
computed by averaging over 20 \AA\ windows centered on 1450 \AA, 2200
\AA, and 4000 \AA.  Finally, we use Equation (\ref{eq:slope}) to
calculate $\alpha$ from 1450-2200 \AA\ at high redshift and from
2200-4000 \AA\ at low redshifts.  We use the uncertainties on flux to
compute the uncertainty in the slope $\sigma_\alpha$, excluding sources
with $\sigma_\alpha > 0.25$.  This predominantly eliminates sources
near the flux limit with poor statistics, reducing the samples to 3646
and 2706 for low and high redshift, respectively.  Since these
QSOs account for only a small fraction of the sample, their exclusion
has little effect on the overall distribution of $\alpha$.

\begin{figure*}
\plotone{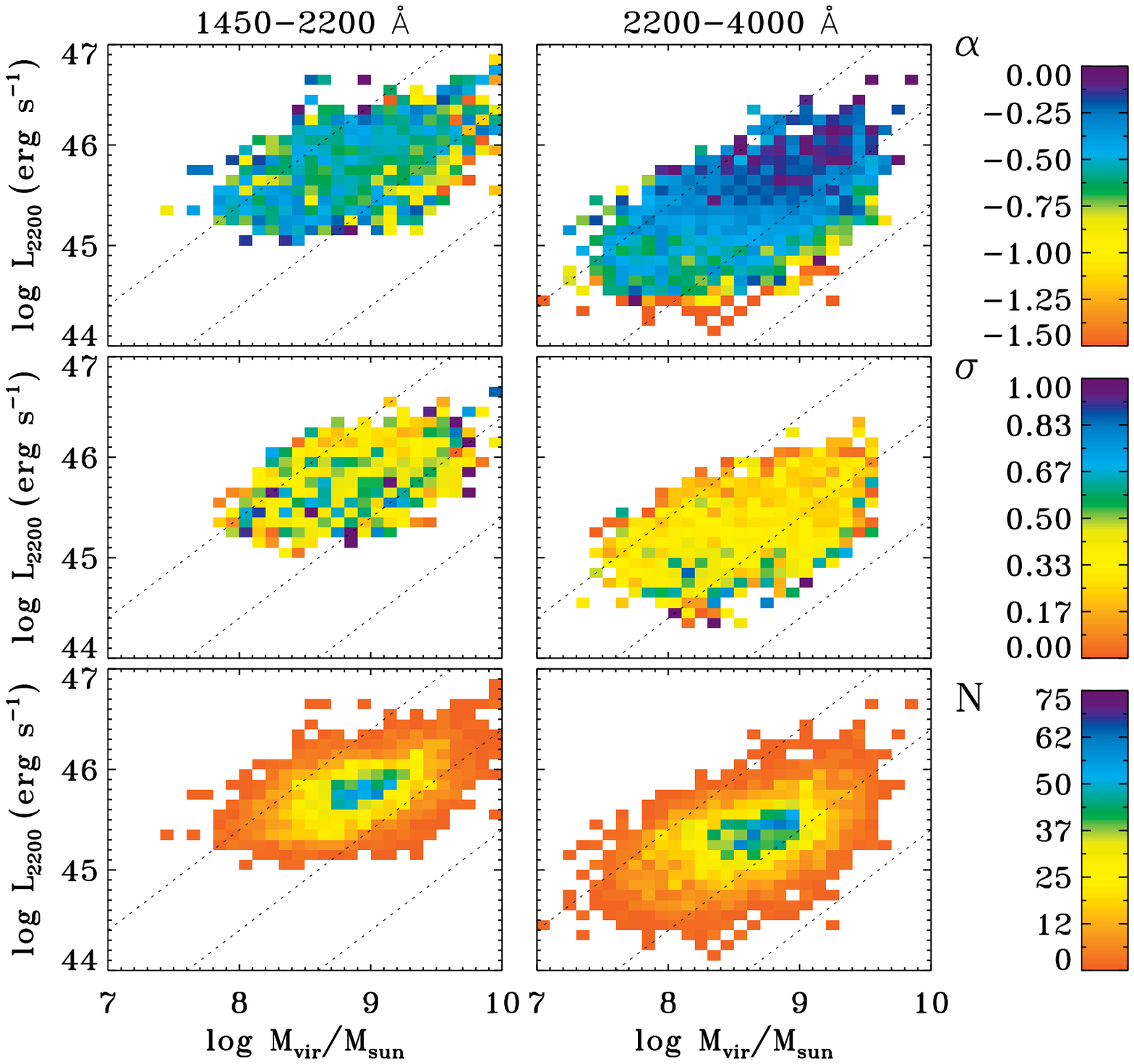}

\caption{2D distributions binned by monochromatic luminosity $L_{2200}$
and the viral mass estimate $\mvir$.  The left and right panels
correspond to the samples with slopes measured from 1450-2200 \AA\ and
2200-4000 \AA, respectively.  The top panels display the mean slopes
$\alpha$, the middle panels show the standard deviation of $\alpha$ in
each bin, and the bottom panels represent the number of spectra
contributing to each bin.  The diagonal dotted lines are approximate
estimates for curves of constant Eddington ratio, assuming $L_{\rm bol} =
5 L_{2200}$. See the text in \S\ref{slopes} for further disscussion.
\label{fig:mass}}
\end{figure*}

The distributions of $\alpha$ for the two samples are plotted in
Figure \ref{fig:slopedist}.  The mean slopes for the two samples are
-0.59 from 1450-2200 \AA\ (thin curve) and -0.37 from 2200-4000 \AA\
(thick curve).  The differences in the distribution of $\alpha$
between the two samples suggests that the QSO spectra may be slightly
concave with redder slopes at shorter wavelengths.  This observation
is consistent with the slope measurement for the relatively nearby AGN
summarized in Table \ref{tbl1}, which are always redder at shorter
wavelengths.

In Figure \ref{fig:1d} we plot the mean slope $\alpha$, binned in
terms of $\mvir$ and $L_{2200} \equiv \lambda L_\lambda$ evaluated at
2200 \AA.  The error bars represent uncertainty estimates for
$\alpha$, corresponding to $\sigma/\sqrt{N}$, where $N$ is the number
of spectra which contribute to the bin and $\sigma^2=\sum
(\alpha-\bar{\alpha})^2/(N-1)$ is the variance in each bin.
Differences in the distributions of $\alpha$ between the two samples
are clearly evident.  The higher redshift, shorter wavelength sample
shows weak anticorrelations of $\alpha$ with $L_{2200}$ and $\mvir$.
In contrast, the lower redshift, longer wavelength sample suggests
that $\alpha$ is positively correlated with both $L_{2200}$ and
$\mvir$.  The strongest trend is the nearly monotonic rise in $\alpha$
as $L_{2200}$ increases.  There also appears to be a weaker trend in
which $\alpha$ increases as $\mvir$ increases to $\sim 10^9 \msun$,
but then turns over and begins to decreases at higher masses.

In order to examine possible correlations between the parameters, we
plot a 2D distribution of $\alpha$ in the top panel of Figure
\ref{fig:mass}.  We bin the data in both $L_{2200}$ and $\mvir$
simultaneously. In this plot and several to follow, the left and right
hand columns correspond to quantities measured using the 1450-2200
\AA\ and 2200-4000 \AA\ samples, respectively.  In the middle panel we
plot the standard deviation $\sigma$, and in the bottom panel we plot
$N$.  The diagonal dashed lines provide a simple, although relatively
crude, estimate of the Eddington ratio.  They are curves of constant
$L_{2200}/\mvir$ and would correspond to lines of constant Eddington
ratio if a linear relationship between bolometric luminosity $L_{\rm
bol}$ and $L_{2200}$ held for all sources.  From right-to-left each
curve represents a factor of 10 increase in this ratio and the
left-most line would correspond to the Eddington limit.  For this
estimate, we assume a ratio $L_{\rm bol}/L_{2200}= 5$, but do not
regard this exact value as particularly significant.  Since the
mean ratio of $L_{2200}/L_{3000} \sim 6/5$ in our sample, this choice
provides approximate consistency with the $L_{\rm bol}/L_{3000} =
5.9$ relation estimated by \citet{mad04}.

We find that the 1450-2200 \AA\ slopes are not strongly correlated
with either $L_{2200}$ or $\mvir$.  The highest concentration of blue
slopes are found at low $L_{2200}$ and low $\mvir$ while the highest
concentration of red slopes occurs at low to moderate $L_{2200}$ and
high $\mvir$ (i.e. low $L/L_{\rm Edd}$). There is a clear trend in the
2200-4000 \AA\ slopes in which $\alpha$ decreases as $L_{2200}$
decreases at fixed $\mvir$.  A weaker variation with $\mvir$ can also
be inferred, even at fixed luminosity, although it is strongest at
high and low $\mvir$ where $N$ tends to be lower.  At fixed
$L_{2200}$, $\alpha$ first increases with increasing $\mvir$ and then
decrease at high $\mvir$ and low $L/L_{\rm Edd}$, consistent with
Figure \ref{fig:1d}. Comparison of $\sigma$ and $\alpha$ in the top
two panels shows that the typical variance in each bin can be quite
large relative to the observed trends.  The strongest trend in the
data, the variation in $\alpha$ from $L_{2200} \sim 5 \times 10^{44} -
10^{46}$ erg s$^{-1}$ at long wavelengths, corresponds to $\Delta
\alpha \sim 0.45$ while $\sigma \gtrsim 0.3$ is common.  We attribute
some, but not all, of this scatter to errors in the flux measurements,
for which $\sigma_\alpha \lesssim 0.15$ is typical.

\begin{figure}
\includegraphics[width=0.46\textwidth]{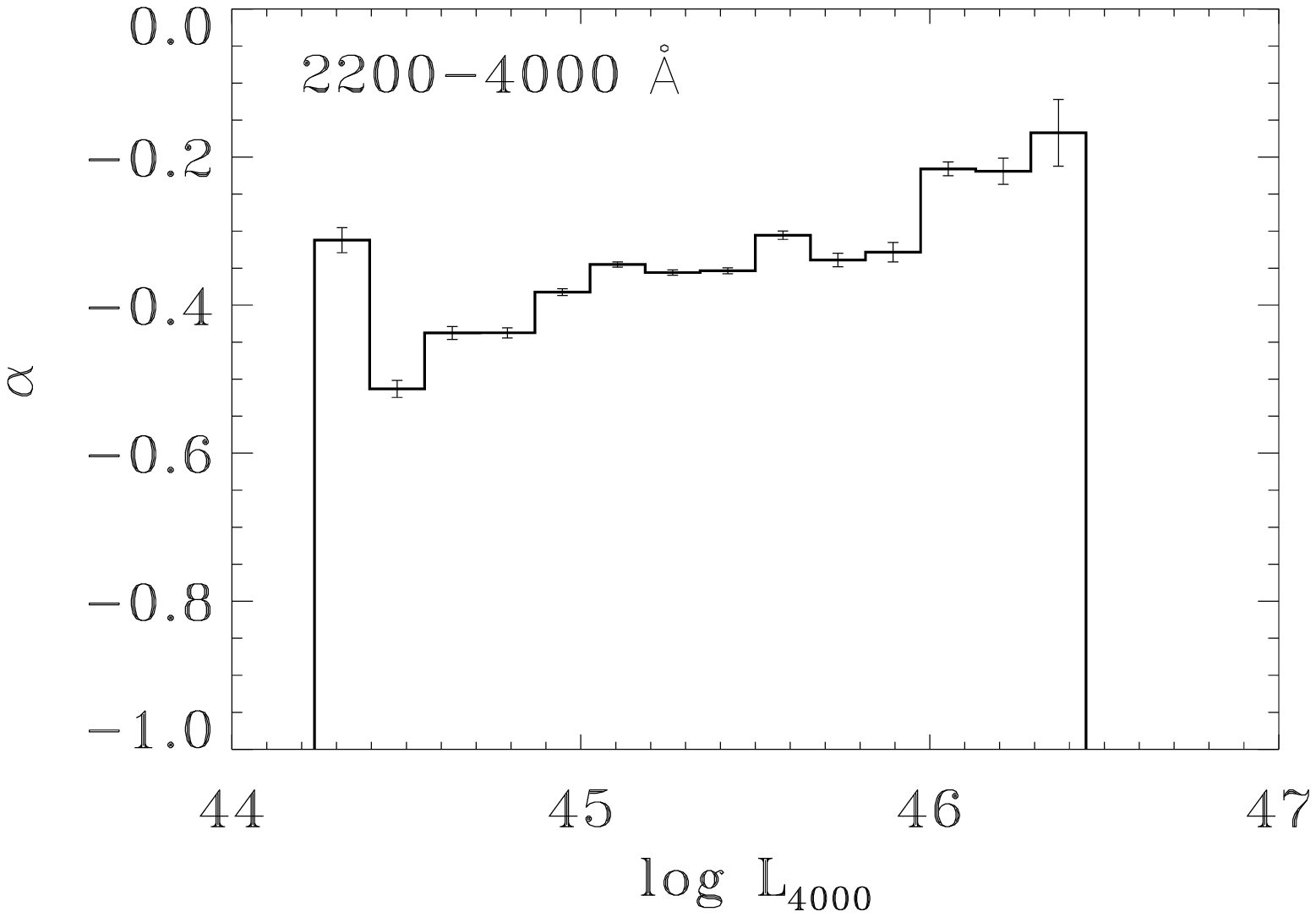}
\caption{1D distribution of mean slopes $\alpha$ binned by
monochromatic luminosity at 4000 \AA.  The slopes are
measured from 2200-4000 \AA. Error bars are calculated assuming
uncertainties $\sigma/\sqrt{N}$ where $\sigma$ and $N$ are the
standard deviation and number of spectra in each bin.
\label{fig:l4000}}
\end{figure}

Obviously, the variation of $\alpha$ with monochromatic luminosity
depends to some extent on the wavelength used to evaluate it. Since it
is the ratio of $L_{2200}$ to $L_{4000}$ determines $\alpha$ in the
first place, a positive (negative) correlation between $\alpha$ and
$L_{2200}$ ($L_{4000}$) would occur if the logarithmic ratio of these
luminosities were randomly distributed about some mean ratio. We plot
the distribution of $\alpha$, binned by $L_{4000}$ in Figure
\ref{fig:l4000}.  As might be expected, we find a weaker, but still
positive, correlation between $\alpha$ and $L_{4000}$ than we found
between $\alpha$ and $L_{2200}$ (cf. Figure \ref{fig:1d}).

\section{Model Slopes and Monte Carlo Comparisons}
\label{mcmod}

\begin{figure*}
\plotone{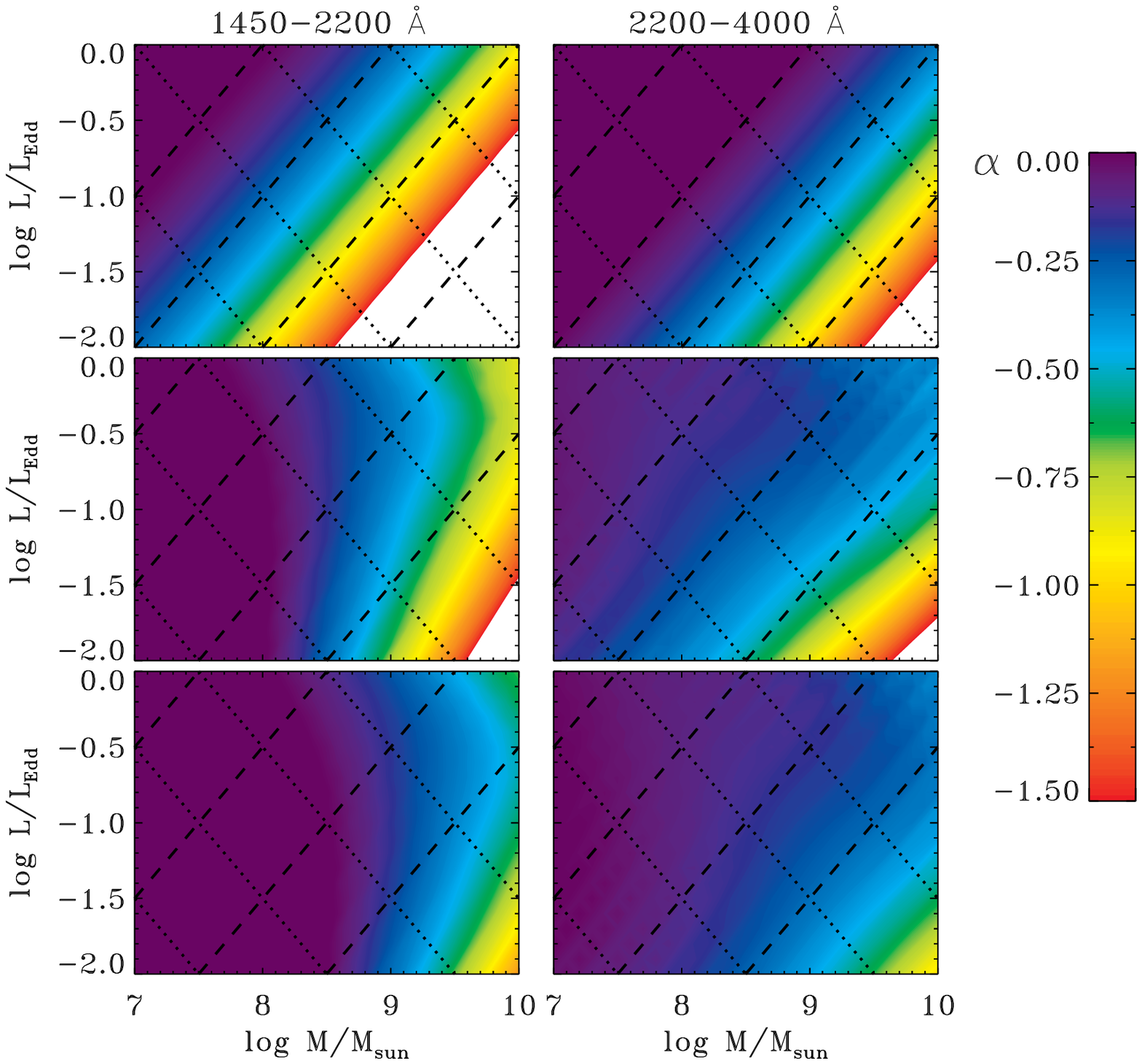}

\caption{Slopes calculated from artificial SEDs of relativistic thin
disk models \citep{hub00}. The left and right panels correspond to the
slopes measured from 1450-2200 \AA\ and 2200-4000 \AA, respectively.
In the upper and middle panels we plot spectra from relativistic
models of accretion onto a Schwarzschild black hole.  The bottom panels
show slopes calculated for a black hole with $a_\ast=0.9$.  For the
spectra in the top panels, we approximate the emission at the disk
surface as blackbody with the local effective temperature.  In the
middle and bottom panels, the emission is based on non-LTE spectral
atmosphere calculations.  The diagonal dashed lines correspond to 
lines of constant $\tin$ and the diagonal dotted lines represent lines
of constant $L$.  See the text in \S\ref{modslopes} for further
discussion.
\label{fig:model}}
\end{figure*}

\subsection{Model Slopes}
\label{modslopes}

We now wish to more carefully examine the model SEDs described in
\S\ref{models}.  Although the models we use are rather sophisticated
in how they treat radiative transfer and disk structure, it is useful
to begin by considering a simple case.  Perhaps the simplest spectral
model one can construct is the multitemperature blackbody.  In this
case one simply calculates the radiative flux in the disk as a
function of radius and computes a spectrum by integrating the emission
over the disk surface, assuming a Planck spectrum $B_\nu$ at the local
effective temperature $\teff$.  This yields
\be
L_\nu=4 \pi^2 \int B_\nu(\teff(R)) R dR.
\label{eq:mtbb}
\ee

To proceed further, one must obtain $\teff$ as a function of $R$.
A common approximation is to assume a power law form for the flux
$F \propto R^{-\beta}$, yielding
\be
\teff=\tin \left(\frac{R}{R_{\rm in}}\right)^{-\beta/4}. \label{eq:teff}
\ee
Inserting this form into equation (\ref{eq:mtbb}) yields
\be
L_\nu=\frac{240 R_{\rm in}^2 \sigma \tin^3 h}{\pi^3 \kb \beta}
\left(\frac{h \nu}{\kb \tin} \right)^{3-8/\beta}
\int_{x_{\rm in}}^{x_{\rm out}} \frac{x^{8/\beta-1}}{e^x-1} dx,
\label{eq:beta}
\ee
where $x\equiv h \nu/ ( \kb T)$.  Note that in addition to the
explicitly power law dependence on $\nu$, $L_\nu$ is also a function of
$\nu$ through the limits of the integral $x_{\rm in} = h \nu/(\kb
\tin)$ and $x_{\rm out} = h \nu/( \kb T_{\rm out})$, For $h \nu \ll
\kb T_{\rm out}$ and $h \nu \gg \kb \tin$ we have $L_\nu \propto
\nu^2$ and $L_\nu \propto \nu^3 \exp{(-h \nu/\kb \tin)}$,
respectively.  At intermediate frequencies $\kb \tin \gg h \nu \gg \kb
T_{\rm out}$, the integral is almost independent of $\nu$ and we find
$L_\nu \propto \nu^{3-8/\beta}$, or $\alpha = 3-8/\beta$.

For a Newtonian thin disk the flux is given by \citep{sas73}
\be
F=\frac{3 G M \mdot}{8 \pi R^3} \mathcal{I},
\label{eq:flux}
\ee 
where $R$ is the radius and $\mathcal{I}$ is a correction factor which
depends on assumptions about the torque at the inner edge of the disk.
Typically, $\mathcal{I}$ is only a weak function of $R$ which
approaches unity at large $R$, so we will ignore it for this simple
example.  Then $\beta \simeq 3$ and $\alpha \simeq 1/3$ well below the
peak in the SED.

For a black hole of mass $M$, it is useful to scale $R$ with $R_g
\equiv G M/c^2$ and $\mdot$ with $\mdot_{\rm Edd}\equiv L_{\rm Edd}/c^2
= 4 \pi G M/(c
\kappa_{\rm es})$, where $\kappa_{\rm es}$ is the electron scattering
opacity.  With the scalings $\dot{m} \equiv \mdot/\mdot_{\rm Edd}$ and
$r=R/R_g$, we find 
\be
\tin =\left( \frac{3 c^5 \dot{m}}{2
\sigma G M \kappa_{\rm es} r_{\rm in}^3}\right)^{1/4}.
\label{eq:tin}
\ee

These results suggest that $\alpha \sim 1/3$ for $h \nu \ll \kb \tin$,
with $\tin$ given by equation (\ref{eq:tin}).  This result, first
derived by \citet{lyn69}, is commonly referenced as a characteristic
spectral slope for accretion disks \citep{fkr92}, but we shall see
below that more sophisticated models generically give lower values of
$\alpha$ for the masses and luminosities considered in this work.  In
part, this will be due to the breakdown of the underlying assumptions:
that the emission is blackbody and that the flux is given by a simple
power law with $\beta = 3$.  However, it will also not always be the
case that $h \nu \ll \kb \tin$ at the UV wavelengths we are
considering.  As $M$ increases or $\dot{m}$ decreases, the frequency
at which the SED peaks ($\nu \sim 3 \kb \tin/h$) decreases.  For $h
\nu \lesssim \kb \tin$, the spectrum should begin to flatten and
$\alpha$ is expected to decrease.

\begin{figure*}
\plotone{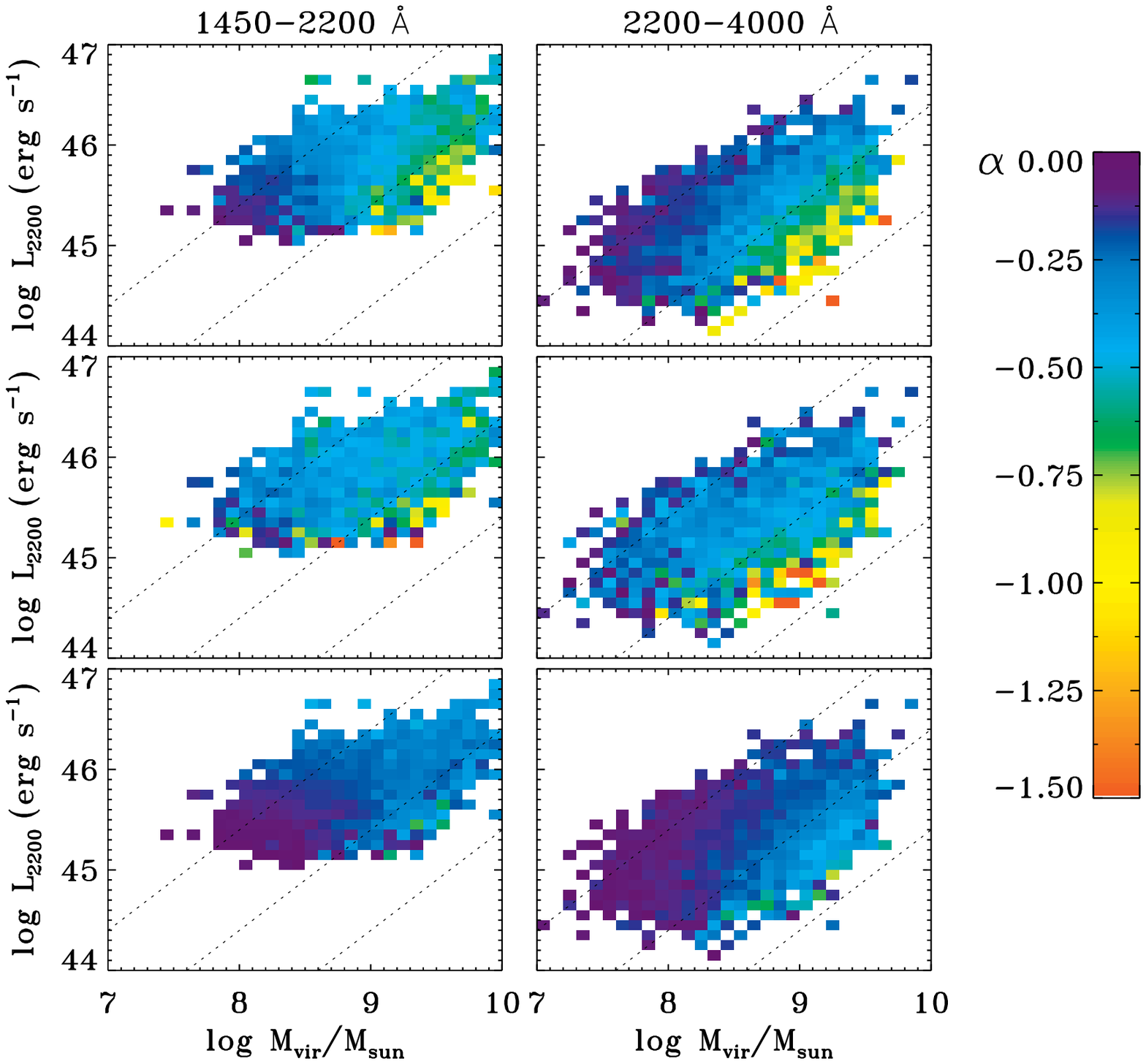}

\caption{Simulated 2D distributions of $\alpha$ binned by $L_{2200}$ and
$\mvir$ for comparison with the top panels of Figure \ref{fig:mass}.
The left and right panels correspond to the slopes measured
from 1450-2200 \AA\ and 2200-4000 \AA, respectively.  The distributions
of $\alpha$ are calculated from thin disk models using a Monte Carlo
method described in \S\ref{mc} to approximate the distribution of 
mass and luminosity.
Each row corresponds to a different
set of assumptions for the black hole spin and the uncertainties in
the mass estimates. In the top and middle panels we plot $\alpha$
calculated for models with $a_\ast=0$ while the bottom panels show the
results for $a_\ast=0.9$.  The top and bottom panels assume $\sigma_M
=0.4$ dex, while the middle panels assume $\sigma_M = 0.8$ dex.
\label{fig:nodust}}
\end{figure*}

This trend is clearly seen in Figure \ref{fig:model}, where we have
calculated $\alpha$ from 1450-2200 \AA\ and 2200-4000 \AA, using
equation (\ref{eq:slope}).  In the top panel, we compute $\alpha$ for
relativistic, multitemperature models with $a_\ast=0$.  We plot
$\alpha$ as a function $L/L_{\rm Edd}$ ($\propto \dot{m}$), where $L$
is the model bolometric luminosity.  The shape of the spectrum, and
therefore, $\alpha$ depends (weakly) on inclination.  Here we plot the
mean $\alpha$, averaged over a uniform distribution in $\cos i$ from
$\cos i = 0.5$ to 1. The diagonal dashed and dotted curves correspond
to lines of constant $\tin$ and constant $L$, respectively.  $L$
increases from bottom left to top right and $\tin$ increases from the
top left to bottom right. In this case, $\alpha$ can be entirely
parameterized by $\tin$.  Since the 1450-2200 \AA\ slopes are measured
with continuum windows nearer to the peak in the SED, it is always the
case that $\alpha$ is lower than measurements made from 2200-4000 \AA.
We note that even for these blackbody models, $\alpha$ is less than
the ``canonical'' value of 1/3 over the entire parameter space of
interest.

In the middle panels of Figure \ref{fig:model} we again plot $\alpha$
for $a_\ast=0$ models, but with spectra based on the atmosphere models
described in \S\ref{models}.  For 2200-4000 \AA\ slopes, we again find
that the variations in $\alpha$ can largely parameterized by $\tin$,
although not completely.  The primary difference is that the slopes
are lower than the multitemperature blackbody models and the overall
variation with $\tin$ tends to be weaker.  There are several effects
which may contribute to these differences, but at these wavelengths
the most important modification to the spectrum is the Balmer edge at
3650 \AA.  For the majority of annuli which contribute to the emission
near 4000 \AA, there is a strong Balmer edge in absorption.  The flux
emitted by any annulus is fixed by the local dissipation of energy, so
the decrement in flux shortward of the edge is compensated by an
increase longward of the edge.  At 2200 \AA\ the effects of the Balmer
edge are minimal.  The net effect is to increase the ratio 4000 \AA\ to
2200 \AA\ flux relative to what it would be if the edge was absent,
yielding a lower $\alpha$.

The differences in the 1450-2200 \AA\ slopes between the two models
are greater.  The contours of constant $\alpha$ are generally more
vertical.  At high $M$ and $L/L_{\rm Edd}$, they actually turn over
and nearly follow the lines of constant $L$ (dotted curves).  Compared
with the multitemperature blackbody models, we find that $\alpha$ is
generally higher, particularly at low $L$, and only at high $L$ is
$\alpha$ lower than the blackbody prediction.  

The differences between the model slopes at shorter wavelengths are
largely due to the increasing importance of electron scattering
opacity in annuli where these photons are primarily emitted.  As the
ratio of scattering to absorption opacity increases the spectrum of a
particular annulus becomes a ``modified blackbody'' \citep[see
e.g.][]{ral79} which peaks at higher photon energies.  Since total
flux is conserved, the increase in higher energy photons must be
offset by a decrease in the flux at lower energies.  The emission at
any particular wavelength $\lambda$ (in this case 1450 \AA) is going
to come from a range of annuli with $T_{\rm eff} \sim c h/(3 \kb
\lambda)$.  When compared with blackbody emission, some of these
annuli will contribute more flux to $\lambda$ and some will contribute
less.  The net effect when emission is integrated over all annuli will
depend on a number of factors: how close $T_{\rm eff}$ is to $\tin$,
the wavelength at which $\kappa_{\rm es}$ becomes the dominant
opacity, and how strongly the spectrum is deformed from blackbody.
For low values of $\tin$ and low $L$ this deformation tends to enhance
the flux at 1450 \AA, but has little effect on the flux at 2200 \AA,
increasing $\alpha$.

In the bottom panels, we again plot slopes for atmosphere based
spectra, but with $a_\ast=0.9$ instead of $a_\ast=0$.  The dependence
of $\alpha$ on $L/L_{\rm Edd}$ and $M$ is very similar to the
$a_\ast=0$ case in that contours of constant $\alpha$ have similar
shapes.  However, the spectra are generically bluer since $\alpha$ is
larger than the $a_\ast=0$ slopes over the whole range covered by the
plot.  There are multiple ways in which changing $a_\ast$ modifies
$\alpha$, but the dominant effect is the shift in the peak of the
SED. In these models, the inner radius $r_{\rm
in}$ is determined by the location of the innermost stable circular
orbit (ISCO), which is smaller for larger $a_\ast$.  From equation
(\ref{eq:tin}), we can see that this makes $\tin$ larger.  At fixed
$L/L_{\rm Edd}$, $\dot{m}$ must also decrease to offset the increase
in radiative efficiency, which would reduce $\tin$.  However, the
former effect dominates, and the peak of the SED shifts to shorter
wavelengths, generally increasing $\alpha$. 

\subsection{Monte Carlo Comparisons}
\label{mc}

With accurate mass, spin, and inclination estimates, we could compare
the slopes in Figure \ref{fig:mass} directly with the models, such as
those shown in Figure \ref{fig:model}.  However, we do not have any
practical means of estimating either the spin or the inclination.
Furthermore, we do not know with certainty the accuracy of mass
estimates $\mvir$ obtained with equation (\ref{eq:mass}).  In order
to account for these uncertainties we construct  slope distributions
similar to those shown in Figure \ref{fig:mass} using Monte Carlo
methods.

In order to characterize the distributions of model parameters we must
make several assumptions.  First we must adopt a distribution of
inclinations.  In absence of obscuration, a uniform distribution in
$\cos i$ would be expected from isotropic emission.  However, the disk
models do not produce isotropic emission.  Disks viewed nearly face on
($\cos i \sim 1$) have larger fluxes than edge-on disks, and should be
somewhat enhanced near the flux limit in a flux limited sample.
Furthermore, angle dependent obscuration is a fundamental tenet of the
unified model of AGN \citep{ant93}.  In the unified model, obscuring
material lies in the disk plane (the ``torus'').  Broad emission line
objects, such as the type I QSOs discussed here, will be viewed nearly
face on, up to the opening angle of the torus.  Here, we adopt a
uniform distribution in $\cos i$ from $\cos i = 0.5$ to 1, where the
lower limit corresponds to an opening angle of $60^\circ$.  Since
the differences in $\cos i$ produce at most a factor of two in flux,
the assumption of a uniform distribution will only produce a small
error near the flux limit.

We must also specify $a_\ast$.  Unfortunately the distribution of
$a_\ast$ is highly uncertain.  Estimates for $a_\ast$ in AGN are
limited to a handful of bright, relatively nearby sources with broad
Fe K$\alpha$ lines.  In some cases, such as MCG -6-30-15, very high
values of $a_\ast$ ($\sim 1$) are inferred \citep[see e.g.][]{tan95}.
A combination of empirical and theoretical arguments seem to favor
$a_\ast \sim 0.9$ \citep{gsm04}, although there are considerable
uncertainties.  Given the large uncertainties, we simply choose two
characteristic spins, $a_\ast=0$ and 0.9, as representative examples.

\begin{figure*}
\plotone{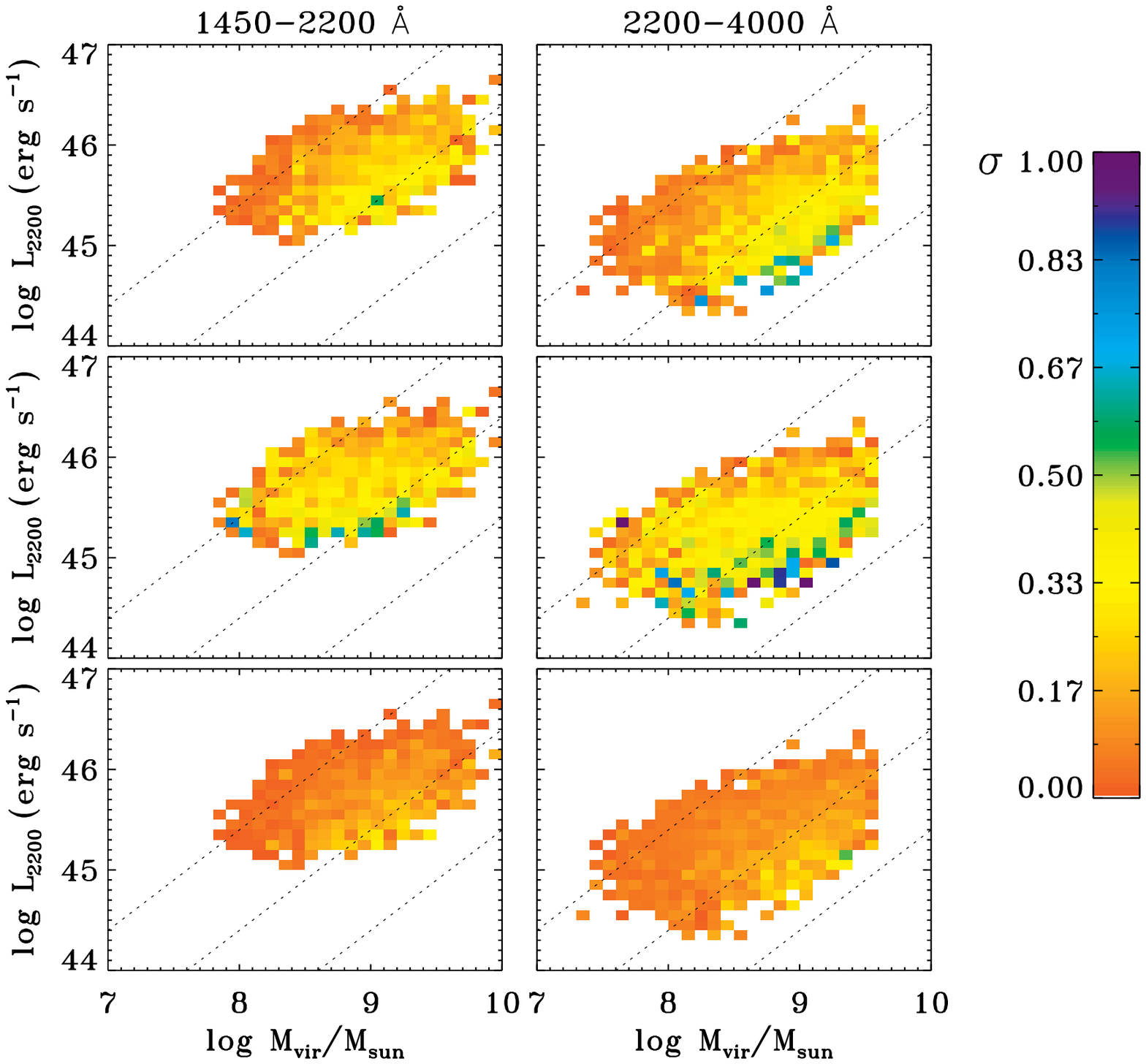}

\caption{The standard deviation $\sigma$ for the simulated
distributions of $\alpha$ plotted in Figure \ref{fig:nodust}.  Each
row correspond to a different set of assumptions for the black hole
spin and the uncertainties in the mass estimates, as described in the
caption of Figure \ref{fig:nodust}.
\label{fig:sigma}}
\end{figure*}

Finally, we must account for the uncertainties in $\mvir$.  Although
these uncertainties are difficult to estimate robustly, some insight
may be obtained by comparing estimates which utilize the H$\beta$ line
width and monochromatic luminosity at 5100 \AA. \citet{mad04} compute
the ratio $\log( \mvir[{\rm H}\beta]/\mvir[{\rm Mg~II}])$, finding a
dispersion of 0.33 dex.  This, of course, does not account for any
systematic errors common to the two methods. The $\mvir[{\rm H}\beta]$
estimates rely on luminosity based $R_{\rm BLR}$ estimates which are
calibrated by matching reverberation mapping estimates. For example,
\citet{ves02} find that 70\% of $\mvir[{\rm H}\beta]$ mass estimates
match the reverberation mapping estimates to within a factor of 3.
The reverberation mapping estimates, in turn, are claimed to have a
typical precision of $\sim 30\%$ \citep{pet04}.  Here, we adopt 0.4
dex as a fiducial value of the uncertainty in $\log \mvir$.  Given the
scatter in the relations discussed above, we view this as a lower
limit on the typical error.  Therefore, we also consider the impact of
assuming a larger uncertainty (0.8 dex) below.

With these assumptions we can begin computing Monte Carlo slope
distributions.  We start by assuming a distributions of $L_{2200}$,
$\vmg$, and $\mvir$ identical to those in \S\ref{slopes}.  Next, we
use random deviates to draw values of the ``actual mass'' $M_0$ and
$\cos i$.  Here, $M_0$ is used to account for possible errors in the
mass estimates.  It is drawn from a log normal distribution with a mean
of $\mvir$ and $\sigma_M=0.4$ dex. This prescription means that the
mass distribution of the models will be somewhat broader than the
distribution of $\mvir$.  However, there is some evidence for a real
cutoff in the mass distribution at high mass.  Therefore, we enforce a
maximum mass $\log M_0/\msun < 9.5$.  Then, for each choice of $M_0$
and $i$, a value of $L/L_{\rm Edd}$ is chosen so the model
monochromatic luminosity matches $L_{2200}$.  Finally, given these
values of $L/L_{\rm Edd}$, $M_0$, $i$, and a choice of $a_\ast$ we can
compute $\alpha$ for the corresponding model. These slopes are then
binned in exactly the same manner as the data, allowing us to compare
with the observed $\alpha$ and $\sigma$. (Note that the distributions
of $N$ are identical to the observations by construction.)

We plot the resulting 2-D distribution of $\alpha$ for $a_\ast=0$ in
the top panels of Figure \ref{fig:nodust}. For both long and short
wavelength slopes, there are clear mass and luminosity dependent
trends in $\alpha$.  At fixed $L_{2200}$, $\alpha$ decreases with
increasing $\mvir$ and at fixed $\mvir$, $\alpha$ generally decreases
with decreasing $L_{2200}$.  Since $L_{2200}$ is roughly proportional
to the bolometric luminosity, the trends are as expected from middle
left panel of Figure \ref{fig:model}.  Comparison with Figure
\ref{fig:mass} shows that these strong variations are clearly
discrepant with the weak trends in $\alpha$ inferred from the
observations.  Although we do not plot the results, we have performed
the equivalent exercise using the multitemperature blackbody slopes.
The model trends are even stronger, due the rapid reddening of the
spectra at high masses and low luminosities.  The discrepancies are
particularly large at short wavelengths, where the multitemperature
blackbody slopes are significantly redder than the observed spectra.

In the the top panel of Figure \ref{fig:sigma}, we plot the
distribution of standard deviations corresponding to the slopes in
Figure \ref{fig:nodust}.  The distributions of $\sigma$ are largely
determined by the assumed uncertainty in $\mvir$ .  If the assumed
uncertainty $\sigma_{\rm M}$ was identically zero, each
$\alpha$ would be almost completely determined by $L_{2200}$ and
$\mvir$ since the variation in $\alpha$ with $\cos i$ is relatively
weak.  In that case, $\sigma \lesssim 0.05$ would be typical.
However, comparison with the middle panel of Figure \ref{fig:mass}
shows that despite this scatter induced by the mass uncertainties,
$\sigma$ for the model slopes is still substantially below the
standard deviations obtained from the observed slopes.

Given the nature of the discrepancy between models and data, an
obvious concern is the effect of errors in the mass estimates.  If the
errors in the mass estimates are larger, they might smear out any mass
dependence in the observed slopes and increase $\sigma$.  Of course,
the actual form of the mass error distribution remains an
important source of uncertainty, and the log normal distribution
employed here may not be an adequate approximation.  However, since we
have no reliable means of independently measuring this distribution,
we simply accept this as a caveat and consider the effects of larger
errors by increasing $\sigma_M$ from 0.4 to 0.8 dex. We plot the
results in the middle panels of Figures \ref{fig:nodust} and
\ref{fig:sigma}.

In the middle panel of Figure \ref{fig:nodust}, we find that the
resulting distributions of $\alpha$ are still incapable of reproducing
the observations, but increasing $\sigma_M$ to 0.8 dex does smear out
the strong variation seen the $\sigma_M=0.4$ dex case, as expected.
At short wavelengths, this removes almost all of the variation in the
slopes with $\mvir$ and $L_{2000}$, in better agreement with the lack
of a strong trend in the observed slopes.  However, the mean slopes
are still larger than those that are observed.  At longer wavelengths,
there is still some trend for larger slopes for higher $L/L_{\rm
Edd}$, which is inconsistent with the observed distribution.  For
$L/L_{\rm Edd} \gtrsim 0.1$, the model slopes are too large at low
$L_{2000}$ and too low at high $L_{2000}$.

In the middle panels of Figure \ref{fig:sigma}, we find that $\sigma$
also increases, as expected.  This provides better agreement with
observed $\sigma$, although $\sigma$ is still generally too low.  We
also find that $\sigma$ decreases as $L/L_{\rm Edd}$ increases, a
trend not seen in the data.  This decrease in $\sigma$ is also
consistent with Figure \ref{fig:model}, which shows that $\alpha$ is a
weaker function of $L/L_{\rm Edd}$ and $M$ for the low $M$ and high
$L/L_{\rm Edd}$ models which predominantly contribute to the low
$\sigma$ bins.

Except at high luminosities and long wavelengths, we nearly always
find that the models predict $\alpha$ larger (bluer) than the
observed values.  A comparison of the bottom and middle panels of
Figure \ref{fig:model} suggests that increasing $a_\ast$ will only
make the discrepancy worse, since the $a_\ast=0.9$ models slopes are
everywhere larger than the $a_\ast=0$ case at equivalent $M$ and
$L/L_{\rm Edd}$.  Indeed, this is precisely what we find in the bottom
panels of Figures \ref{fig:nodust} when we plot $\alpha$ for
$a_\ast=0.9$ and $\sigma_M=0.4$ dex.  For the range of $\mvir$ and
$L_{2000}$ sampled, $\alpha$ varies less strongly than in the
$a_\ast=0$ case.  However, this also means that $\sigma$ is even
lower, increasing the disagreement with the data.  Thus, it appears
that the $a_\ast=0.9$ models are an even poorer match to the data that
the $a_\ast=0$ models. This result may be problematic since, as
discussed above, there is some evidence which suggests $a_\ast \gtrsim
0.9$ may be more common than $a_\ast \sim 0$.

\begin{figure*}
\plotone{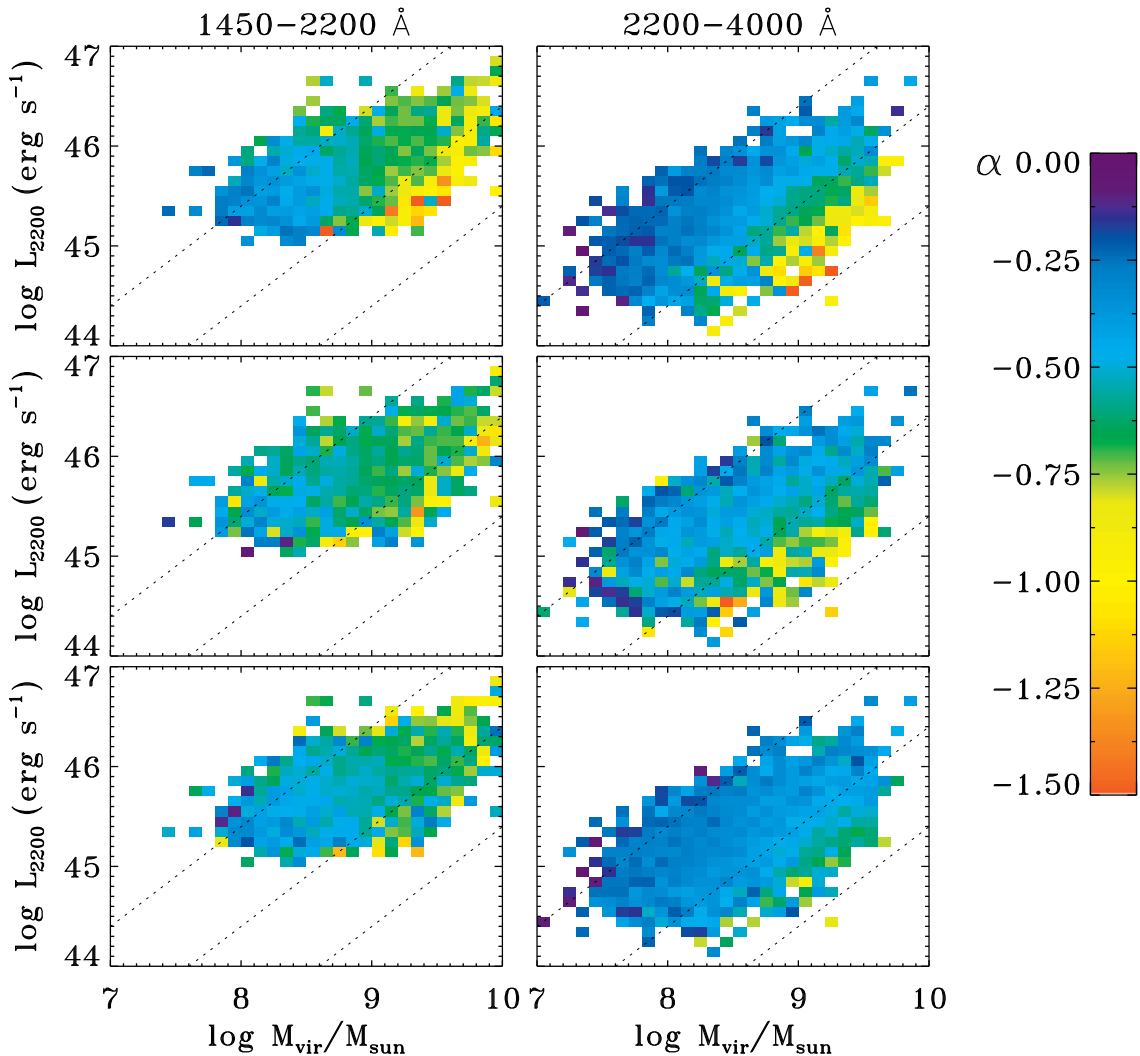}

\caption{Simulated 2D distribution of $\alpha$ binned by $L_{2200}$ and
$\mvir$.The left and right panels correspond to the slopes measured >
from 1450-2200 \AA\ and 2200-4000 \AA, respectively.  The
distributions account for the effects of dust extinction with an
SMC-like reddening curve (see text), but are otherwise equivalent to
those plotted in Figure \ref{fig:nodust}.  We calculated each
distribution assuming a uniform distribution in E(B-V) between 0 and a
specified maximum value chosen to match the observed short wavelength
slopes of Figure \ref{fig:mass}.  We assume a maximum $\rm E(B-V) =
0.03$ for the $a_\ast=0$ models (top and middle panels), and $\rm
E(B-V) = 0.03$ for the $a_\ast = 0.9$ models (bottom panels).
\label{fig:dust}}
\end{figure*}

\section{Discussion and Conclusions}
\label{discus}

\subsection{The Effects of Dust Reddening on Spectral Slopes}
\label{dust}
It is clear that some additional process (or processes) must be
modifying the spectrum if the models are to be reconciled with the
observed slopes and we discus a number of possibilities below.  One of
the main discrepancies is that the observed spectra are generally
redder than the models predict.  One likely possibility is that the
SEDs are altered by wavelength dependent extinction from dust
intrinsic to the source or host galaxy.  Dust emission can be seen in
the infrared in many AGNs, and is almost certainly present at some
level in our sample \citep[see e.g.][]{ric03}.  The uncertainties are
mainly questions of how much dust is present and what is the
wavelength dependence of the extinction.

We can incorporate the effects of dust reddening directly into our
Monte Carlo distributions by simulating the effect of extinction on
the model SEDs.  Multiplying our artificial spectra by a reddening
curve will change both the values of $\alpha$ and $L_{2200}$ inferred
by an observer.  In order to proceed we must specify the shape of the
reddening curve and specify the amount of extinction.  Often, the
degree of reddening is parameterized by the color excess E(B-V) which
is the difference in extinction between the B and V bands, expressed
in magnitudes.  Since this distribution is unknown, we simply adopt a
uniform distribution of color excess between zero and some specified
maximum value.

The wavelength dependence of the extinction is also uncertain. The
determination of QSO reddening curves is an area of active research
and ongoing debate \citep{ric03,gas04,hop04,cze04,wil05}.
Much of the difficulty in determining these reddening curves stems
from the problem of disentangling intrinsic variations in the quasar
SED (lines or continuum) from the effects of dust. This makes it
difficult to unambiguously identify ``unreddened'' AGN sources for
comparison with a ``reddened'' population.  Despite this uncertainty,
it is widely accepted that the $\sim 2200$ \AA\ feature seen in the
Galactic interstellar medium \citep[see e.g.][]{ccm89} is extremely
weak or absent in AGN reddening curves.

Disagreement primarily arises over the steepness of the reddening
curve, particularly at wavelengths shortward of $\sim 3000$ \AA.  For
example, \citet{gas04} derive a flat far UV reddening curve, while the
analysis of \citet{hop04} favors a steeper, SMC-like curve.  Another
calculation by \citet{cze04}, which is based on ratios of the color
selected SDSS quasar composites of \citet{ric03}, prefers a curve
which is in between: somewhat flatter than the SMC, but still steeper
than that found by \citet{gas04}.

Clearly, if extinction is important, the observed slope distributions
will depend significantly on the form of the reddening curve.
Therefore, we have considered three possibilities: the approximate
forms of the curves derived by \citet{gas04} (see their Appendix) and
\citet{cze04} (equation [3]), and the SMC-like curve used by
\citet{ric03} ($A_\lambda=1.39 \; E(B-V) (\lambda/\mu {\rm
m})^{-1.2}$).

In Figure \ref{fig:dust} we plot Monte Carlo distributions equivalent
to those in Figure \ref{fig:nodust}, but including the effects of
dust.  For this example we use the SMC-like reddening curve and choose
maximum E(B-V) so that short wavelength slopes approximately match the
observed slopes (cf. Figure \ref{fig:mass}).  This matching requires
E(B-V) $\sim 0.03$ and $0.055$ for $a_\ast = 0$ and 0.9, respectively.
These modest values would not contradict the results of \citet{ric03}
who found that 94\% of their sample would be consistent with E(B-V) $<
0.04$ for the same reddening curve. We note that the modification of
$L_{2200}$ due to extinction is typically small for the required range
of E(B-V).  In principle, a larger amount of extinction could
introduce a correlation where dust reddened sources are less luminous
than bluer, unreddened spectra, and possibly explain the observed
trend of higher $\alpha$ for higher $L_{2200}$.  Also, the small
amount of extinction at 2200 \AA\ introduces considerably less scatter
in $\alpha$ than the uncertainties in the mass estimate.  Therefore,
we do not replot the standard deviation of the bins, since the plots
differ little from those of Figure \ref{fig:sigma}.

The reddening curve of \citet{cze04} has a flatter wavelength
dependence than the SMC-like curve, and therefore requires greater
extinction to produce the same amount of reddening.  In order to match
the short wavelengths slopes of Figure \ref{fig:mass}, we require a
maximum E(B-V) $\sim0.6$ and 0.12 for $a_\ast=0$ and 0.9,
respectively.  The relative flatness of this reddening curve in
comparison to the SMC-like curve also leads to a greater degree
reddening at longer wavelengths to produce the same amount of
reddening a short wavelengths.  This is particularly significant for
$a_\ast=0.9$ model.  Since the observed slopes are redder at lower
$L_{2200}$, this improves the agreement with the data for low
$L_{2200}$, but provides a poorer match for higher values of
$L_{2200}$, where slopes are bluer.  

The \citet{gas04} curve is nearly flat shortward of 3000 \AA.
Therefore, it requires substantially greater extinction to compensate
for the slope discrepancy between the models and data at short
wavelengths. We cannot easily parameterize the amount of extinction
required because it is large enough to boost the intrinsic
(unreddened) luminosity by an order of magnitude or more.  This would
make the models substantially super-Eddington for the assumed masses,
exceeding the upper limit of our grid where the thin disk assumptions 
are no longer self-consistent.

Although our assumed distribution of color excess which is
uncorrelated with luminosity can improve the agreement between the
models and data at short wavelengths, it cannot reproduce the
$L_{2200}$ dependent trend at longer wavelengths.  To some degree,
this conclusion hinges on our assumption that the amount of dust
extinction is independent of luminosity. If, for example, the amount
of dust extinction anticorrelates with luminosity, the most luminous
QSOs could have unreddened, intrinsically blue spectra while lower
luminosity QSOs would have lower slopes due to the dust reddening, in
agreement with the observed trend.  However, the lack of a similar
luminosity dependent correlation at short wavelengths would then
require a reddening curve which is flat shortward of $\sim 2200$ \AA,
such as that proposed by \citet{gas04}.  Since the short wavelength
slopes are redder than predicted by the models, matching the data with
the models at short wavelengths is not possible with such a flat
reddening curve.  Therefore, even if such luminosity dependent redding
is plausible, it seems unable to simultaneously account for both the
short and long wavelength spectral slopes if the underlying continuum
is well approximated by the models.  However, such a scenario may
account for the observed trend if the short wavelength continuum is
intrinsically redder than the models predict.

\subsection{The Effects of Irradiation}
\label{irrad}

Self-irradiation presents another possibility for explaining the red
slopes in these systems.  In fact, correlated variability on
timescales comparable to the light travel times suggests that
irradiation must be occurring at some level \citep[see e.g.][]{kro91}.
In the UV, we expect irradiation to become increasingly less important
for determining the spectrum as we move to shorter wavelengths in the
UV, since local dissipation {\it must} dominate the flux near the peak
of the SED.  Nevertheless, irradiation may still play some role so we
briefly consider its effects by examining a simple model.

The slope modification due to reddening will be strongly dependent on
the geometry of the accretion flow.  A simple estimate of the slopes
of irradiated disks may be obtained from equation (\ref{eq:beta})
which implies $\alpha=3-8/\beta$ for $h \nu \ll \kb \tin$.  In order to
estimate $\beta$, we need to specify the radial dependence of the
irradiated flux $F_{\rm irr} \propto R^{-\beta}$.  With simple
geometric arguments, one can show \citep[see e.g.][]{bla04a} that 
\be
F_{\rm irr} = \frac{L_\ast(1-a)}{4 \pi R^2}\left(\frac{H}{R} \right)
\left(\frac{d \log H}{d \log R}-1 + \frac{H_\ast}{R} \right). 
\label{eq:firr} 
\ee 
In order to obtain this relation, we approximate the irradiating
continuum (the inner disk, or possibly a corona) as a point source
$L_\ast$ at $R=0$ and a height $H_\ast$ above the midplane.  The disk
surface at the point of irradiation is parameterized by the height
$H(R)$ at radius $R$.  The albedo $a$ may also be a function of $R$,
but we will ignore this dependence for simplicity.

For $H_\ast \gg H(R)$, we find $F_{\rm irr} \propto R^{-3}$.  This
dependence is easily understood by noting that emission from the point
source falls off as $R^{-2}$ and is intercepts the disk surface with
angle $\theta$ such that $\cos \theta = H_\ast/R$.  Since $H_\ast$ is
independent of $R$, we find $F_{\rm irr} \propto R^{-3}$ as inferred
from equation (\ref{eq:firr}). If this reprocessed flux dominates the
locally dissipated flux and is reradiated as a blackbody we again have
$\beta=3$ and $\alpha=1/3$, equivalent to the bare thin disk case.

If $H_\ast \leq H(R)$, the disk must flare in order to produce
significant irradiation.  If we parameterize $H$ as a power law ($H
\propto R^\Gamma$), we find $\beta=3-\Gamma$ or $\alpha=(1-3 \Gamma)/
(3-\Gamma)$.  An important case is $\Gamma = 1$ for which $\beta=2$
and $\alpha=-1$.  For $\Gamma > 1$, $\alpha < -1$ and vice-versa.
This suggests that in the region of parameter space where most of our
slopes lie, $-1 < \alpha < 1/3$, we require $\Gamma < 1$.  Such models
are concave in shape and self-shielding at sufficiently large radii.

Obtaining slopes in this range via irradiation may therefore require
some level of fine tuning.  One possibility is that the disk
transitions from a flat or convex ($\Gamma > 1$) solution to a concave
($\Gamma < 1$) solution at the range of radii which give rise to the
UV emission so that self-shielding occurs only at larger radii and
longer wavelengths.  A second possibility is that there is flaring
with $\Gamma > 1$, but that the reprocessed flux does not dominate the
local dissipation.  The differing fraction and radial dependences of
the local dissipation and reprocessed flux give rise to a range of
$\alpha$ between $\sim -1$ and $\sim 1/3$.  Such a concurrence would
be somewhat surprising, because a comparable contribution from the
locally dissipated and reprocessed emission will only occur over a
limited range of radii, due to their different radial dependences.

Nevertheless, self-shielding geometries may be useful for explaining
some aspects of the observed slope distribution.  In the top panel of
Figure \ref{fig:mass} we see that for the highest luminosity sources,
slopes are, on average, bluer at longer wavelengths than at short
wavelengths.  The redder than expected slopes at short wavelengths
could be produced by irradiation of a portion of the disk surface
which blocks emission from reaching larger radii.  The unirradiated
flow at larger radii and lower $T_{\rm eff}$ could remain dominated by
the local dissipative flux and produce an intrinsically bluer slope.

In order for such interpretations to be valid, the geometry of the
irradiated accretion flows needs to be explained.  Our thin disk
models for supermassive black holes are radiation pressure dominated
in their inner-most radii \citep{sas73}, implying a scale height which
is nearly independent of radius.  With this model, we would not expect
reprocessing to significantly alter the spectrum until the disk
transitions to the gas pressure dominated regime where $H \propto
R^{21/20}$ \citep{sas73}, yielding $\alpha \sim -1$ if reprocessing
dominates the local flux.  The transition radius from radiation to gas
pressure dominance depends on $M$, $\alpha_{\rm SS}$, $\dot{m}$, and
$a_\ast$, but is generally located at 200-400 $R_g$. 

Since the bulk of the UV radiation in our models is radiated at radii
$\lesssim 200 R_g$, some other mechanism must be modifying the disk
structure in order to explain the red slopes via irradiation. One
possibility is that the vertical extent of these disks may be
substantially modified due to the magnetic support, as suggested by
numerical simulations \citep[see e.g.][]{tur04,hks06}.  However, at
present, such calculations remain too uncertain to yield a predictive
model for the reprocessing.  Another possibility is that backscattered
radiation from an outflow might modify the spectrum
\citep[e.g.][]{nik04}, but this will depend on the (unknown) outflow
geometry, even if such outflows prove to be common.

\subsection{The Dependence of Slope on Luminosity}
\label{lumdep}

As discussed in \S\ref{mc} and \S\ref{dust}, the luminosity dependent
slopes at long wavelengths present a challenge for the \citet{hub00}
models since the discrepancies cannot be simply attributed dust
reddening or errors in the mass estimates.  As discussed in
\S\ref{dust}, adding an ad-hoc luminosity dependence to the dust
extinction might account for the dependence of $\alpha$ on $L_{2200}$
at long wavelengths, but not without creating problems at shorter
wavelengths.

Due to the short baselines involved ($\log (4000/2200) = 0.26$ and
$\log (2200/1450) = 0.18$) the slopes can be substantially modified by
relatively small changes in flux of a continuum window.  Therefore, we
consider ``contamination'' from the host galaxy and/or the BLR
emission as a possible solution.  In such a scenario, the observed
trend would imply either an increasing contribution to flux at 4000
\AA\ as $L_{2200}$ decreases or an increasing contribution to the flux
at 2200 \AA\ as $L_{2200}$ increases.  Inspection of the bottom, left
panel of Figure \ref{fig:1d} shows a variation in the mean slope $\Delta
\alpha \sim 0.45$ from $L_{2200} < 10^{45} \; \rm erg \; s^{-1}$ to
$L_{2200} \gtrsim 10^{46} \; \rm erg \; s^{-1}$.  This is only
slightly greater than the typical standard deviation in the individual
bins and corresponds to a $\sim 30\%$ variation in the relative flux
between the two continuum windows.

One scenario would be a contribution from the host galaxy which is
roughly independent of the QSO luminosity and contributes $\sim 30\%$
at low $L_{2200}$.  Based upon the strength of absorption lines in
their composite spectrum \citet{van01} estimate a 7\%-15\%
contribution at the location of Ca~II $\lambda 3933$ and Na~I $\lambda
5896$.  Therefore, the trend cannot be explained entirely by host
galaxy contamination.  A second possibility is excess BLR emission at
2200 \AA\ which correlates with $L_{2200}$.  Inspection of the
\citet{van01} composite spectrum indicates emission from high
excitation Fe~II lines is probably the main contaminant at $\sim 2200$
\AA.  However, a comparison of composite spectra for QSOs with
$L_{2200}$ above and below $3 \times 10^{45} \; \rm erg s^{-1}$ does
not provide evidence for such a large change in the equivalent width of
these lines so we conclude that this explanation is also unlikely.

It is also possible that the underlying accretion flow is thin disk,
but that the \citet{hub00} models have not properly accounted for the
emission near $\sim 4000$ \AA.  The model grid only extends
to $\teff = 10^4 \; \rm K$, so the emission for annuli in the disk at
larger radius and lower $\teff$ are approximated by blackbodies.  For
models with $\teff \lesssim 10^4 \; \rm K$ density inversions can
occur due to a fall in electron pressure when H recombines \citep[see
Fig. 9 of][]{hub00}.  This makes it very difficult to predict the true
equilibrium structure, and therefore the spectrum, of an annulus in a 
real, turbulent accretion flow.

The lack of models $\teff < 10^4 \; \rm K$ creates a problem because
these annuli may still have significant Balmer edges.  The presence of
edge at $\sim 3650$ \AA\ tends to create an excess of emission longward
of the edge and a deficit at shorter wavelengths relative to the
predictions of a pure blackbody.  As can be seen by Figure 11 of
\citet{hub00}, replacing non-LTE model spectra with blackbodies may
produce a reduction in the flux at 4000 \AA.  This result, coupled
with the observation that the strength of the edge is anticorrelated
with luminosity at fixed mass \citep[see Figure 13 of][]{hub00},
suggests that the Balmer edge may play a greater role in producing the
observed trend than the current models predict.  A robust
determination of the spectrum at these wavelengths might ultimately
require spectral models coupled with realistic disk simulations, and
is beyond the scope of this work.

It is also possible that selection biases and incompleteness play some
role in producing the observed trends.  The spectroscopic targeting of
QSOs in SDSS is discussed in \citet{ric02}.  For this sample, we
consider incompleteness near the flux limit to be the greatest cause
for concern since the majority of our sources are well separated from
the stellar locus.

The spectroscopic targeting algorithm rejects sources based upon $i$
band magnitude.  For $ugri$ and $griz$ selected sources the magnitude
limits are $i \lesssim 19.1$ and $i \lesssim 20.2$, respectively.  The
$i$ band filter is approximately centered at $\sim 7600$ \AA, which
corresponds to $\sim 3800$ \AA and $\sim 2500$ \AA for the low ($z
\sim 1$) and high ($z \lesssim 2$) redshift samples respectively.
This could lead to a deficit of blue quasars at low $L_{2200}$ in the
long wavelength, low redshift sample. For an equivalent $L_{2200}$,
red quasars will have a larger flux near 4000 \AA\ than sources with
bluer slopes, and would be more likely to make it into the
spectroscopic sample if they are near the $i$ flux limit.  We have
considered this possibility by plotting the slopes as function of $i$
for our sample, but we do not find a significant deficit of blue
slopes near the flux limits.  In fact, we find a decreasing fraction
of red quasars for $i \lesssim 17.5$.  This deficit of red spectra
with large flux in $i$ seems to be due to a real paucity of red sources
with high luminosities.

\begin{figure*}
\plotone{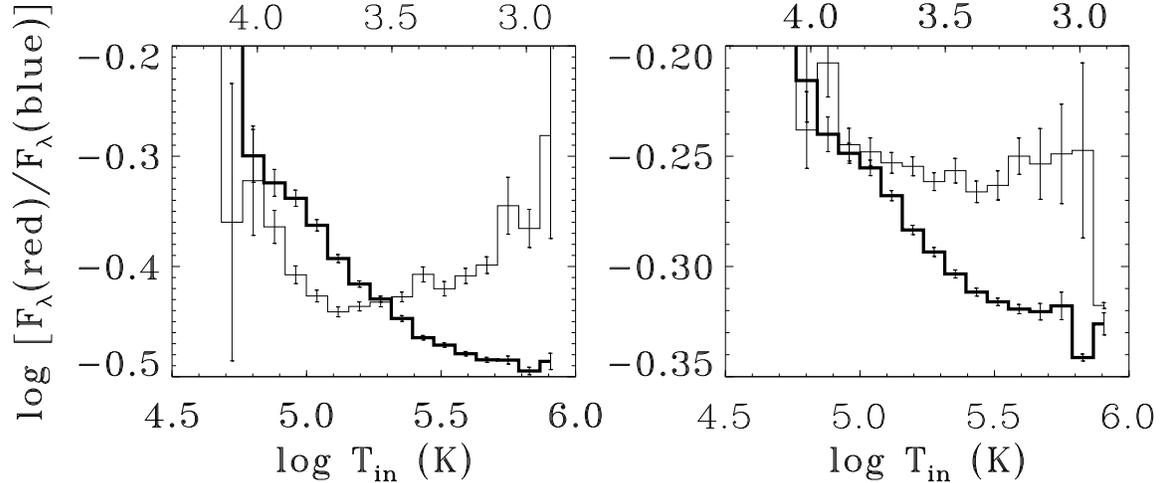}

\caption{Distribution of flux ratios binned as a function of $\tin$ for
comparison with the results of \citet{bon06}. The right and left
panels correspond to the slopes measured from 1450-2200 \AA\ and
2200-4000 \AA, respectively.  We calculate $\tin$ from $\vmg$ using
the relation $\log \tin = 5.43 - \log(\vmg/3000 \; {\rm km \;
s^{-1}})$.  The top axis displays the value of $\log \vmg$ for
comparison with Figure \ref{fig:1d}. We plot the distribution of the
observed SDSS flux ratios as a thin curve.  The solid curves show
distributions of model flux ratios based on the Monte Carlo
distributions.  The plotted curves correspond to the same distribution
plotted as a 2D distribution in the top panel of Figure \ref{fig:nodust}
with $a_\ast=0$ and $\sigma_M=0.4$ dex.
\label{fig:colors}}
\end{figure*}

\subsection{Comparison with Previous Work}
\label{comp}

There have been a number of previous observational tests of the
\citet{hub00} models considered here.  For the most part, these
comparisons have involved individual sources or a relatively small
sample with broadband spectral coverage. \citet{bla01} fit models to
spectra of 3C 273, finding poor agreement in the near UV.  The
reverberation mapping mass estimate has since been revised, bringing
the model and data into better, though not perfect agreement
(cf. \S\ref{fuse}).

\citet{bla04b} calculated spectral slopes as a function of mass and
Eddington ratio for a subset of sources in the \citet{sha05} sample.
The results showed significant scatter and rather poor agreement with
the model predictions, but the analysis only included a few dozen
objects.  Using the same data, \citet{sha05} compared the optical,
optical-to-UV, and far UV spectral slopes to the models, finding rough
agreement. A recent analysis of SDSS quasars observed by the {\it
Galaxy Evolution Explorer} (GALEX) \citep{tra07} also examined the far
UV properties of AGNs, again finding evidence for slope changes near
$\sim 1000$ \AA, in approximate agreement with the model predictions.

A recent analysis by \citet{bon06} relates most directly to our
current work. They use another, overlapping sample of SDSS
QSO spectra and measure color ratios between continuum windows located
at 1350 \AA, 2200 \AA, 4000 \AA, and 5100 \AA.  They also compare
their results with the a selection of \citet{hub00} models chosen to
approximate the distribution of Eddington ratios and masses inferred
from the data.

The choice of variables used for binning is one of the major
differences between the \cite{bon06} analysis and this
work. \citet{bon06} consider the evolution of observed and artificial
spectra as a function of a single variable corresponding to $\tin$.
For the viral mass estimates used here, the dependence of $\tin$ on
$\vmg$ and $L_{3000}$ can be obtained by combining equations
(\ref{eq:mass}) and (\ref{eq:tin}).  If we make the additional
assumption that $L_{\rm bol} \propto L_{3000}$, we find 
\bd 
\tin \propto L_{3000}^{(1-2 \delta)/4} \vmg^{-1}.  
\ed 
Following \citet{mad04}, we have used $\delta=0.62$ which yields a
weak dependence on luminosity.  \citet{bon06} use $\delta=0.5$ exactly
so that $\tin$ is a function of $\vmg$ alone. 

To facilitate comparison, we have replotted our distributions as a
function of $\tin$ in Figure \ref{fig:colors}, using flux ratios in
place of $\alpha$ on the vertical axis.  Following equation (5) of
\citet{bon06}, we use the relation $\log \tin = 5.43 - \log(\vmg/3000
\; {\rm km \; s^{-1}})$ for the horizontal axis.  We plot the
distributions of both the observed (thin curve) and model (thick
curve) flux ratios.  The model curves are calculated using the same
Monte Carlo distribution as in the top panel of Figure
\ref{fig:nodust}.  At long wavelengths the observed SEDs are reddest
at low and high $\tin$.  This differs from the model which are red at
low $\tin$ and become monotonically bluer as $\tin$ increases.  At
short wavelengths the observed flux ratios are roughly independent of
$\tin$, while the model fluxes are again reddest at low $\tin$ and
become monotonically bluer as $\tin$ increases.  These results are
qualitatively consistent with Figure 3 of \citet{bon06}, who measure
colors at 1350 \AA\ as opposed to the 1450 \AA\ window used here.

\citet{bon06} infer that the observed reddening at high $\tin$ may be
related to Eddington ratio since most of the objects contributing to
the highest $\tin$ bins have $L/L_{\rm Edd} \gtrsim 0.3$.  This is
also roughly consistent with our findings.  In our sample this result
can be understood by examining the top left panel of Figure
\ref{fig:mass} and considering equation (\ref{eq:tin}).  Figure
\ref{fig:mass} shows that most of the QSOs in our sample do not
radiate significantly above the Eddington limit.  Equation
(\ref{eq:tin}) implies that the highest values of $\tin$ are obtained
for low values of $M$ and high Eddington ratios (i.e. high $\dot{m}$).
As a result, the sources in our sample with the highest $\tin$ (and,
therefore, low $\vmg$) tend to occupy the low $\mvir$, low $L_{2200}$
corner of the plot.  Since the observed slopes are predominant
functions of $L_{2200}$ with redder slopes at lower luminosities,
these QSOs also tend to be redder than average.  Of course, bins with
slight lower $\tin$ also include low $L_{2200}$ objects, but the
average $\alpha$ is still larger due to the increasing fraction of
higher $\mvir$ and $L_{2200}$ sources, which tend to be bluer.
Therefore, the problem of understanding the redding at low $\tin$ is
intimately connected to the question of why the mean slopes are
predominantly functions of $L_{2200}$ (rather than $L_{2200}/\mvir$) at
long wavelengths which we discussed in \S\ref{lumdep}:

\subsection{Conclusions}
\label{conc}

We have shown how slopes from artificial SEDs of thin accretion disks
\citep{hub00} vary with black hole mass and bolometric disk
luminosity.  As expected from naive models, we find that the slope
$\alpha$ generally decreases as $M$ increases at fixed $L$.  We have
shown that the UV spectral slopes of models based on radiative
transfer calculations differ measurably from those with simple
blackbodies, and considered how the slopes are modified by changes in
the black hole spin.

We first compared the models against five broadband SEDs of nearby,
bright AGN \citep{sha05} which also have reverberation mapping mass
estimates. For the more luminous sources, the models can roughly
reproduce the observed flux ratios in continuum windows at 1450 \AA,
2200 \AA, and 4000 \AA.  At lower luminosities, the short wavelength
slopes are substantially redder than the model predictions, and may
indicate significant reddening by dust local to the source or host
galaxy.  We then measured $\alpha$ for 6352 QSOs, using these same
continuum windows.  We find only a weak trend with mass when virial
estimates are used at short (1450-2200 \AA) or long (2200-4000 \AA)
wavelengths.  Even if we allow for errors in the mass estimates with a
log normal distribution and a standard deviation of 0.4 dex, a much
stronger mass dependent trend is observed in the model slope
distributions which is not consistent with the observations.

A possible explanation is that the mass estimates typically have
larger errors than our assumed distribution predicts. In support of
this possibility, we find that increasing the typical
mass estimate error  to $\sim 0.8$ dex is sufficient to erase
most of the mass dependent trend in the model slopes.  However,
this increase alone is insufficient to bring the slope distribution
into agreement with the data.

The multitemperature blackbody models yield slopes which are too red
at short wavelengths for black holes with $M \gtrsim 10^9 \msun$.
With the exception of the longer wavelength slopes of the most
luminous QSOs, we always find that the observed slopes are redder than
the non-LTE atmosphere based model predictions for Schwarzschild black
holes.  The discrepancy is even greater for spinning black holes which
are generally bluer than their non-spinning counterparts.  We suggest
that much of this discrepancy could be accounted for by dust reddening
in the source or host galaxy.  If we use an SMC-like reddening curve
\citep{ric03}, we require E(B-V) $\lesssim 0.03$ and 0.055 in order to
obtain agreement between the short wavelength slopes.  Reddening
curves which are flatter at wavelengths shortward of 2200 \AA\
\citep[e.g.][]{cze04,gas04} would require a greater color excess.  If
the reddening curve is as flat as that derived by \citet{gas04}, dust
reddening will have little impact on the short wavelength slopes, and
the observed slopes should very nearly match the intrinsic slopes.  In
that case, the models considered here would not be consistent with the
observed slopes.

At longer wavelengths (2200-4000 \AA), the observed slopes are
generally bluer at high $L_{2200}$ and redder at lower luminosities,
for a fixed ratio relative to the Eddington luminosity.  The models,
however, predict a trend which is predominantly determined by
Eddington ratio and are not consistent with this result.  This
discrepancy remains even after we simulate the effects of dust
reddening.  This discrepancy may partly arise from difficulties in
properly modelling emission near the Balmer edge which can affect
the flux at 4000 \AA\ significantly.  Improving the models at
these (and longer) wavelengths is an important step for future
studies.

Overall, we find no clear signature of bare, thin accretion disks from
the distribution of observed UV spectral slopes.  Nevertheless, we do
not believe the present analysis is sufficient to rule out a dominant
contribution from such models.  We consider uncertainties in the
amount of extinction, wavelength dependence of reddening curve, and
precision of the mass estimates to be the most important caveats.  If
the mass estimates prove to be sufficiently precise (i.e. correct to
within a factor of four) or dust reddening proves sufficiently weak,
the actual accretion flows must differ significantly from the models
employed here.

\acknowledgements{We thank S. Antonucci, E. Bonning, I. Hubeny, S. Jester, 
J. Kollmeier, R. Lupton, G. Shields, and N. Zakamska for useful 
discussions.  We are grateful to R. McLure and J. Dunlop for making their 
table of Mg~II line width measurements available to us and Z. Shang et al. 
for making their broadband spectra publicly available. This work was 
supported by NSF grant AST03-07657 and NASA grant number PF6-70045 awarded 
by the Chandra X-ray Center, which is operated by the Smithsonian 
Astrophysical Observatory for NASA under contract NAS8-03060.}

\end{document}